\documentstyle[12pt]{article}
\def\afour{n}    
\if\afour y
    \def\fileversion{1.2}
    \def\filedate{26 Feb 90}
    
    \advance\textheight by \topskip
\else
\fi
\newcommand{\sect}[1]{\section{#1}\setcounter{equation}{0}}
\newcommand{\subsec}{\subsection}
\newcommand{\ds}{\vrule height 8mm width 0pt \displaystyle}
\newcommand{\bbbr}{\mbox{\rm I}\!\mbox{\rm R}}
\newcommand{\ie}{{i.e.}}
\newcommand{\eg}{{e.g.}}
\newcommand{\zhs}{\hskip 0pt plus 0pt}
\newcommand{\hy}{-\penalty10000\zhs}
\newcommand{\Et}{--\penalty10000\zhs}
\newcommand{\beq}{\begin{equation}}
\newcommand{\eeq}{\end{equation}}
\newcommand{\md}{\mbox{\rm d}}
\newcommand{\D}{\mbox{\rm D}}
\newcommand{\E}{\mbox{\rm E}}
\newcommand{\G}{\mbox{\rm G}}
\newcommand{\Dq}{\bar{\mbox{\rm D}}}
\newcommand{\e}{\mbox{\rm e}}
\newcommand{\imu}{\mbox{\rm i}}
\newcommand{\pdf}{\partial}
\newcommand{\abs}[1]{\left\vert#1\right\vert}
\newcommand{\lwc}[1]{\left.#1\right\vert}
\newcommand{\kr}{{\cal y}}
\newcommand{\ihalf}{{\textstyle {\mbox{{\sevenrm i}}\over 2}}}
\newcommand{\half}{\mbox{${1\over 2}$}}
\newcommand{\quart}{\mbox{${1\over 4}$}}
\newcommand{\eighth}{\mbox{${1\over 8}$}}
\newcommand{\sixteenth}{\mbox{${1\over 16}$}}
\newcommand{\isixteenth}{{\textstyle {\mbox{{\sevenrm i}}\over 16}}}
\newcommand{\can}{\mbox{\sevenrm can}}
\newcommand{\Tr}{\mbox{\rm Tr}}

\newcommand{\s}{\mbox{\rm s}}
\newcommand{\q}{\mbox{\rm q}}
\newcommand{\qq}{\bar{\q}}
\newcommand{\vev}[1]{\left<#1\right>}

\begin{document}
\font\csc=cmcsc10 scaled \magstep2
\font\sevenrm=cmr7 scaled \magstep1
\font\twelverm=cmr12
\title{\hfill {\twelverm SI--95--4\\
             \hfill hep-th/9506162}\\
\if\afour y\vrule height 2.5cm width 0cm
\else\vrule height 0.7cm width 0cm
\fi
{\bf Topological Yang--Mills Theory\\ with Two Fermionic Charges.\\
     A Superfield Approach\\ on K\"ahler Manifolds}}
\author{{\csc
        H.\ D.\ Dahmen,
        S.\ Marculescu\thanks{E-mail:\quad{\tt
            marculescu@hrz.uni-siegen.d400.de}}} \ and {\csc
        T.\ Portmann\thanks{E-mail:\quad{\tt
            portmann@sicip1.physik.uni-siegen.de}}}\and
        {\em Universit\"at\hy Gesamthochschule\hy Siegen,\/}\and
        {\em D--57068 Siegen,
            Germany\/}}
\date{(to be published in Nucl.\ Phys.\ B)}
\maketitle
\begin{abstract}
The four\hy dimensional topological Yang\Et Mills theory with two
anticommuting charges is naturally formulated on K\"{a}hler manifolds. By
using a superspace approach we clarify the structure of the Faddeev\Et Popov
sector and determine the total action. This enables us to perform perturbation
theory around any given instanton configuration by manifestly maintaining all
the symmetries of the topological theory. The superspace formulation is very
useful for recognizing a trivial observable (\ie\ having vanishing correlation
functions only) as the highest component of a gauge invariant superfield.  As
an example of non\hy trivial observables we construct the complete solution to
the simultaneous cohomology problem of both fermionic charges.  We also show
how this solution has to be used in order to make Donaldson's interpretation
possible.
\end{abstract}
\newpage
\renewcommand{\arraystretch}{1.4}
\sect{Introduction}
\label{intro}
There is a renewed interest in topological Yang\Et Mills (TYM) theory \cite{I}
over the past two years. While an immediate physical meaning is still a matter
of debate (see however \cite{II}, \cite{an}), TYM certainly offers powerful methods
\cite{III}, \cite{Mo} for extracting non\hy perturbative information 
\cite{IV}, \cite{V} about supersymmetric chromodynamics.

Before presenting the content of our paper we would like to give a short
review of TYM. For more details one should consult the excellent articles
\cite{VII}, \cite{VIII}.

\subsec{Review of Topological Yang\Et Mills Theory}
\label{review}
The basic property of TYM is that its action can be written as the
variation of some gauge invariant expression. The variation itself acts on
the gauge field as a shift and is nilpotent. The former property allows one to
formulate TYM on curved manifolds.

There are essentially two ways of deriving the TYM action, each of them having
its own merit. One can, for instance, construct an action incorporating the
following set of subsidiary conditions: self\hy duality (instanton), fixing of
the topological shift and the BRS gauge fixing \cite{IX}, \cite{X}. Being
directly related to instanton calculus it can be conveniently used for explicit
calculations \cite{XI}, \cite{Anse}. 

Another way \cite{I} to obtain TYM is by twisting the Euclidean $N=2$
supersymmetric gauge theory and by coupling thereafter to external gravity.
While the last step breaks down the original $N=2$ supersymmetry, some
supersymmetries may exist on the curved background as global symmetries. We
call them fermionic symmetries.

A single fermionic symmetry is preserved on arbitrary Riemannian manifolds. One
can show~\cite{XIII} that two or four symmetries remain unbroken if the
manifold is K\"{a}hler or hyper\hy K\"{a}hler, respectively, because each
K\"ahler structure is equivalent to a corresponding Killing spinor.

For the rest of this short review the TYM has a single fermionic symmetry
$\q$, hence it is formulated on a Riemannian four\hy manifold. The action
produced by twisting differs by $\q$\hy exact 
terms from that obtained via subsidiary conditions. However, the correlation
functions are the same \cite{XI}.

The TYM obtained by twisting gains in clarity when formulated in superspace
\cite{XIV} despite the consequence that BRS symmetry has to be introduced in
terms of superfields.

TYM imposes very strong restrictions on the possible observables. Only those
objects which are not highest components of gauge invariant superfields can
have non\hy trivial correlations. The proper mathematical background for
constructing such observables is equivariant cohomology \cite{Cartan}. Its use
in TYM has been initiated by \cite{Ouvry} and further developped in
\cite{Kalkman}, \cite{Stora}. 

The only known examples of TYM observables are the Donaldson polynomials
\cite{Don}, (for a review see \cite{DK}).
They can be obtained \cite{Brooks}, \cite{Dah} by twisting some components of
the $N=2$
superconformal anomaly \cite{Sohn}. By an appropriate change of renormalization
prescriptions one can obtain the Donaldson polynomials in the semiclassical
approximation. However, this approximation turns out to be sufficient as a
consequence of the non\hy renormalization theorem for chiral fields
\cite{Gris}, \cite{Dunbar}.

Further restrictions on the observables come through the path integral
representation. For TYM there is a canonical functional measure \cite{I} $[\md
m][\md\widetilde m]$, where $m$ denotes the moduli and ${\widetilde m}$
their $\q$\hy transformation. The above measure is supplemented with simple
prescriptions for integrating out all non\hy zero modes \cite{I} (as well as
some
zero modes \cite{Anse}).  The integration over the fermionic zero modes amounts
to replacing ${\widetilde m}$ everywhere by $\md m$, such that the correlation
function becomes the integral of a certain top\hy form over the moduli space
\cite{I}.

Another consequence of the fermionic zero modes ${\widetilde m}$ is the
vanishing of the partition function. This means that the action of TYM has an
Abelian global invariance---hereafter called ghost number symmetry---that is
not shared by the path integral. The amount ${\Delta}U$ of violation of the
ghost number symmetry is obtained by integrating the twisted $\bbbr$\hy anomaly
\cite{Dah}, \cite{Wu} over the manifold. One can choose the ghost number $U$ 
such that its
variation ${\Delta}U$ exactly compensates the dimension of the moduli space of
self\hy dual instantons \cite{Atiy}. This provides us with the following
selection rule:
The total ghost number of the observables in a non\hy vanishing correlation
function equals the dimension of the instanton moduli space.

\subsec{Content of the Paper}
\label{content}
In the present paper we consider TYM with two anticommutig symmetries along 
with the items described previously. The twisting appropriate for this case has
been found for the first time in \cite{XIII}. Due to the existence of a Killing
spinor which is necessary to preserve the second supersymmetry the Riemannian
manifold has an integrable complex structure whose K\"ahler form is closed,
thus being in fact a K\"{a}hler manifold. 

We use this formulation in section~\ref{super} to set up a superspace
approach. By extending the method of \cite{XIV} to two Grassmann variables we
encounter constraints in superspace. They are solved in terms of a
prepotential, a pair of chiral\hy antichiral connections and a chiral\hy
antichiral
pair of antisymmetric tensors. As a consequence, the gauge symmetry is
replaced by the local chiral symmetry. Furthermore, the gauge symmetry is
completely fixed if one is postulating subsidiary conditions for the chiral
connections. The gauge and Faddeev\Et Popov terms are then easily constructed.

Section~\ref{symme} deals with the symmetries of the theory. On K\"{a}hler
manifolds each conservation law involves two contragredient (with respect to
conservation indices) but inequivalent tensors. They belong to different
superfields and have different positions within each multiplet. Most prominent
examples are the energy\hy momentum tensor and the pair of fermionic symmetry
currents. These objects represent (up to improvement terms which leave the
conservation law unchanged) the highest components of gauge invariant
superfields.

The `classical' theory under discussion is invariant under two global Abelian
symmetries, whereas one encounters anomalies at the quantum level. In 
section~\ref{pertu} we use this opportunity to check our superspace formalism in
perturbation theory. We evaluate the gravitational contribution to both
Abelian anomalies and found that they are equal. Moreover, they are equal to
the dimension of the instanton moduli space. The result coincides with the one 
obtained for a Riemannian manifold admitting a K\"{a}hler structure.

In section~\ref{donal} we give the solution to the simultaneous 
cohomology problem of both fermionic charges. 
In contrast to TYM with only one fermionic symmetry, this solution consists of 
local observables which may
depend on a larger number of fields, including antisymmetric fields.
We can, however, show that the correlation functions of the integrated 
observables can be interpreted as integrals of top-forms over the moduli 
space of selfdual instantons. 

A different treatment of TYM with two fermionic charges has been given in
\cite{Park} and \cite{Holo}. A short comparision with our work is provided 
at the end of section~\ref{donal}.

\sect{Superspace Approach}
\label{super}
Consider a superspace with two Grassmann coordinates $\theta$,
$\bar\theta$ related by complex conjugation $\imu\bar\theta =
\overline{\imu\theta}$. The commuting coordinates $z^m$  and $z^{\bar m}$
with $m = 1,2$, $\bar m = \bar 1, \bar 2 $ represent local holomorphic and
antiholomorphic coordinates, respectively on a K\"{a}hler manifold
${\cal K}$ without boundary and endowed with the metric $(1,1)$\hy form
\beq\label{i}
\gamma = g_{m \bar n}\md z^m \md z^{\bar n} = \pdf\bar\pdf h \; .
\eeq
Here, $h$ is the K\"{a}hler potential of the metric.

To these coordinates one associates the superconnections
$A_{m}$, $A_{\bar m} = - (A_{m})^\kr$, $A_{\theta}$ and
$A_{\bar{\theta}} = - (A_{\theta})^\kr$, in short $A_M$ with $M = m, \bar m,
\theta, \bar{\theta}$. Under an infinitesimal gauge transformation they
change as
\beq\label{ii}
\delta A_M = \D_M K = {\nabla}_M + [ A_M, K ] \;
\eeq
where $K = - K^\kr$. We use ${\nabla}_m$, ${\nabla}_{\bar m}$ to represent
K\"{a}hler derivatives, while $\theta$, $\bar{\theta}$ directions are flat,
\ie
${\nabla}_{\theta} = {\pdf}_{\theta}$, ${\nabla}_{\bar{\theta}} =
{\pdf}_{\bar{\theta}}$.

We also introduce the following covariant superfields: the anti\hy Hermitean
scalar $\Lambda = - {\Lambda}^\kr$ and the pair of complex conjugate,
antisymmetric and anticommuting tensors $X_{mn}$,
$X_{{\bar m}{\bar n}} = - (X_{mn})^\kr$. They transform as
\beq\label{iii}
\delta X = [ K, X ] \;
\eeq
where $X = \Lambda, X_{mn}$ or $X_{\bar m \bar n}$.

All superfields are taken in an irreducible representation of the compact and 
semi\hy simple group
$\G$ with anti\hy Hermitean generators $t_a = - t^\kr_a$ where $a =
1,\ldots, n=\dim \G$ and with totally antisymmetric structure constants 
$c_{abc}$ defined
through $[ t_a, t_b ] = c_{abc} t_c$. The generators $t_a$ are normalized by
$\Tr (t_a t_b) = - {\delta}_{ab}$.

From superconnections one can construct the superfield strengths $F_{mn}$,
$F_{m{\theta}}$, $F_{m{\bar{\theta}}}$,
$F_{\theta \theta}$, their complex conjugates, $F_{m{\bar n}}$ and
$F_{\theta {\bar{\theta}}}$. They are defined by
\beq\label{iv}
F_{MN} = {\nabla}_M A_N + A_M A_N - (-)^{|M||N|}\left( M \leftrightarrow N
\right) \;
\eeq
where $\abs{M}$ is the grading of $M$, \ie\ $\abs{M} = 0$ for $M = m, {\bar m}$
and $\abs{M} = 1$ for $M = \theta, \bar{\theta}$. The superfield strengths
transform covariantly, \ie\ as~(\ref{iii}).

\subsec{Constraints in Superspace}
\label{constra}
In order to obtain the desired TYM one imposes the constraints
\beq\label{v}
F_{\theta \theta} = F_{\bar{\theta} \bar{\theta}} =
F_{m \bar{\theta}} = F_{\bar{m} \theta} = 0  \; ;
\eeq
\beq\label{vi}
\D_{\bar{\theta}}X_{mn} + i F_{mn} = 0 \; ; \qquad
\D_{\theta}X_{\bar m \bar n} + i F_{\bar m \bar n} = 0 \; .
\eeq
Notice the similarity of~(\ref{v}) with the constraints in superspace for the
$N=1$ supersymmetric gauge multiplet \cite{Wess}. The role of~(\ref{v}) is to
ensure that TYM has a single gauge field.

The fermionic symmetries are represented by the Grassmann derivatives
$\q=\pdf_\theta$ and $\qq=\pdf_{\bar\theta}$. They are nilpotent and
anticommute: $\q^2=\qq^2=\{\q,\qq\}=0$.

\subsec{Wess\Et Zumino Gauge}
\label{wesszu}
We define the fields of TYM through covariant superfields and their Grassmann
covariant derivatives:
\beq\label{vii}
\begin{array}{rclrcl}
\psi_m &=& -\lwc{F_{m\theta}}\; ;\qquad &
\psi_{\bar m} &=& -\lwc{F_{\bar m\bar\theta}}\; ;\\
\varphi &=& -\imu\lwc{F_{\theta\bar\theta}}\; ;\qquad &
\lambda &=& \lwc{\Lambda}\; ;\\
g_+ &=& 2\lwc{\D_\theta \Lambda}\; ;\qquad &
g_- &=& 2\lwc{\D_{\bar\theta}\Lambda}\; ;\\
k &=& \lwc{\left[\D_\theta,\D_{\bar\theta}\right] \Lambda}\; ;\qquad &
f_{m\bar n} &=& \lwc{F_{m\bar n}}\; ;\\
\chi_{mn} &=& \lwc{X_{mn}}\; ;\qquad &
\chi_{\bar m\bar n} &=& \lwc{X_{\bar m\bar n}}\; ;\\
b_{mn} &=& \lwc{\imu \D_\theta X_{mn}}\; ;\qquad &
b_{\bar m\bar n} &=& \lwc{\imu \D_{\bar\theta} X_{\bar m\bar n}}\; ;\\
f_{mn} &=& \lwc{\imu \D_{\bar\theta} X_{mn}}\; ;\qquad &
f_{\bar m\bar n} &=& \lwc{\imu \D_\theta X_{\bar m\bar n}}\; .\\
\end{array}
\eeq
The vertical bar means the lowest component of the superfield, \ie\ at
$\theta=\bar\theta=0$.

Further components of these superfields can be obtained from~(\ref{vii}) with
the help of Bianchi identities and constraints~(\ref{v}), (\ref{vi}). They
are related to~(\ref{vii}) through the gauge covariant derivatives
$\D_m$, $\D_{\bar m}$ or vanish.

The transformations of the fields defined by~(\ref{vii}) are generated by the
Grassmann covariant derivatives
\beq\label{viii}
\delta\left(\lwc{X}\right)
= \imu \left( \zeta + \imu \xi \right) \lwc{\D_\theta X}
+ \imu \left( \zeta - \imu \xi \right) \lwc{\D_{\bar\theta} X} \; .
\eeq
Here $X$ represents any of the covariant superfields on the rhs.\
of~(\ref{vii}); $\zeta$ and $\xi$ are real, anticommuting parameters.

From~(\ref{viii}) one can see that the fermionic transformation
$\q=\pdf_\theta$ (or $\qq=\pdf_{\bar\theta}$ resp.) is always accompanied by
a gauge transformation $\lwc{\left[A_\theta,X\right]}$ (or
$\lwc{\left[A_{\bar\theta},X\right]}$ resp.) restoring the gauge covariance.
From the transformation of the various field strength $f_{mn}$, $f_{\bar
m\bar
n}$, and $f_{m\bar n}$ one can deduce how the gauge fields are transforming,
albeit up to a gauge transformation.  Usually, in the Wess\Et Zumino (WZ)
gauge the last gauge transformation is suppressed. A simple calculation leads
to the following transformation rules:
\beq\label{ix}
\vcenter{\hbox{%
$\begin{array}{rclrcl}
\delta a_m &=& \imu(\zeta+\imu\xi)\,\psi_m\; ;\;&
\delta a_{\bar m} &=& \imu(\zeta-\imu\xi)\,\psi_{\bar m}\; ;\\
\delta \psi_m &=& (\zeta-\imu\xi)\,\D_m\varphi\; ;\;&
\delta \psi_{\bar m} &=& (\zeta+\imu\xi)\,\D_{\bar m}\varphi\; ;\\
\delta \varphi &=& 0\; ;\\
\delta \lambda &=& \ihalf(\zeta+\imu\xi)\,g_+ + \ihalf(\zeta-\imu\xi)\,g_-;\\
\delta g_+ &=& -\imu(\zeta-\imu\xi)\,(k-\imu[\varphi,\lambda])\; ;\;&
\delta g_- &=& -\imu(\zeta+\imu\xi)\,(k+\imu[\varphi,\lambda]);\\
\end{array}$}\hbox to 125mm {\hfill
$\begin{array}{rcl}
\delta k &=& \frac{1}{2}(\zeta+\imu\xi)\,[\varphi,g_+]
          -  \frac{1}{2}(\zeta-\imu\xi)\,[\varphi,g_-]\; ;\\
\delta \chi_{mn} &=& (\zeta+\imu\xi)\,b_{mn}
                  +  (\zeta-\imu\xi)\,f_{mn}\; ;\\
\delta \chi_{\bar m\bar n} &=& (\zeta+\imu\xi)\,f_{\bar m\bar n}
                  +  (\zeta-\imu\xi)\,b_{\bar m\bar n}\; ;\\
\delta b_{mn}
     &=& -\imu(\zeta-\imu\xi)\,(\D_m\psi_n-\D_n\psi_m+[\varphi,\chi_{mn}])\;
;\\
\delta b_{\bar m\bar n}
     &=& -\imu(\zeta+\imu\xi)\,(\D_{\bar m}\psi_{\bar n}-\D_{\bar n}\psi_{\bar
m}+[\varphi,\chi_{\bar m\bar n}])\; .\\
\end{array}$\hfill}}
\eeq
This choice of the field components enables one to declare them (K\"ahler)
metric independent. It follows that fermionic symmetries and the variation with
respect to the metric always commute.

The transformations~(\ref{ix}) close into $\varphi$\hy field dependent gauge
transformations.

In section~\ref{donal} we shall give the fermionic transformations generated
by the nilpotent, anticommutative operations $\q$ and $\qq$.

\subsec{Action}
\label{action}
The action in superspace is given by
\beq\label{x}
{\cal S} = \quart \int_{\cal K} \md^2 z \md^2 {\bar z} \; g \;
\pdf_\theta\pdf_{\bar\theta} \; \Tr \left\{ -
\quart X_{mn}X^{mn} + \Lambda\left(\imu F -
                 [\D_\theta,\D_{\bar\theta}]\Lambda\right)\right\}\;
\eeq
where $g =\det g_{m\bar n}$ and $F = g^{\bar nm}F_{m\bar n}$. Indices
are raised by the inverse K\"{a}hler metric $g^{\bar nm}$ of the K\"ahler
manifold ${\cal K}$.

The equations of motion in superspace are
\beq\label{xi}
\begin{array}{rcl}
\D_\theta X_{mn} &=& 0\; ;\qquad\qquad \D_{\bar\theta} X_{\bar m\bar n} = 0\; ;\\
\half\,\D^n X_{mn} + \D_\theta \D_m \Lambda &=& 0\; ;\qquad
\half\,\D^{\bar n} X_{\bar m\bar n} + \D_\theta \D_{\bar m} \Lambda = 0\; ;\\
F+2\imu[\D_\theta,\D_{\bar\theta}]\Lambda &=& 0\; ;\\
\{\D_m,\D^m\}\Lambda + \ihalf\{X_{mn},X^{mn}\}
     &+& 4\imu\{\D_\theta \Lambda,\D_{\bar\theta} \Lambda\}
     - 2\imu\left[\Lambda,[\Lambda,F_{\theta\bar\theta}]\right]=0\; .
\end{array}
\eeq
From~(\ref{x}) one can get the action in component fields:
\beq\label{xii}
\begin{array}{rcl}
{\cal S} &=& \eighth \ds\int_{\cal K} \md^2 z \md^2 {\bar z} \; g \;
\Tr \left\{\lambda\{\D_m,\D^m\}\varphi + \imu fk -k^2
+ \half\left(f_{mn}f^{mn}\right.\right.\\
&& \left. - b_{mn}b^{mn}\right) + \imu\left(g_+\D_m\psi^m + g_-\D^m\psi_m -
\chi_{mn}\D^m\psi^n - \chi^{mn}\D_m\psi_n\right)\\
&& \left. + 2\imu\lambda\{\psi_m,\psi^m\}
+ \varphi\left(\half\{\chi_{mn},\chi^{mn}\} + \{g_+,g_-\}\right) -
[\varphi,\lambda]^2
\right\}\; .
\end{array}
\eeq
It coincides (up to some field redefinition) with that obtained in \cite{XIII}
by
twisting $N=2$ supersymmetric Yang\Et Mills theory (SYM).

\subsec{Solution of the Constraints}
\label{solut}
One cannot solve the constraints~(\ref{vi}) in superspace. Instead one can
take them into account by means of Lagrange multipliers. For this purpose one
introduces a new action
\beq\label{xiii}
\begin{array}{rcl}
{\widetilde{\cal S}} &=& {\cal S} + \sixteenth\ds\int_{\cal K}\md^2 z \md^2
{\bar z}\, g\,\left(\pdf_\theta \Tr \left\{L^{mn}\left(\D_{\bar\theta}X_{mn}
+ \imu F_{mn}\right)\right\}\right.\\
&& + \left.\pdf_{\bar\theta} \Tr
\left\{L_{mn}\left(\D_\theta X^{mn} + \imu F^{mn}\right)\right\}\right) \;
\end{array}
\eeq
where $L_{mn}$ and $L_{{\bar m}{\bar n}} = - (L_{mn})^\kr$ are a pair of
complex conjugate, anticommuting, and antisymmetric superfields satisfying
\beq\label{xiv}
\D_\theta L_{mn} = \D_{\bar\theta} L_{\bar m\bar n}
= 0 \; .
\eeq
The (covariant) superfields entering~(\ref{xiii}) are subject to the
constraints~(\ref{v}) and~(\ref{xiv}).

The solution of these constraints can be given in terms of
\begin{itemize}
\item
a Hermitean prepotential $V=V^\kr$,
\item
a chiral\hy antichiral pair of superconnections $\phi_m$, $\phi_{\bar m} =
-(\phi_m)^\kr$ depending on a single Grassmann variable
$\pdf_{\bar\theta}\phi_m = \pdf_\theta \phi_{\bar m} = 0$, and
\item
a pair of chiral\hy antichiral, anticommuting, and antisymmetric superfields
$M_{mn}$, $M_{\bar m\bar n} = -(M_{mn})^\kr$ obeying $\pdf_\theta M_{mn} =
\pdf_{\bar\theta} M_{\bar m\bar n} = 0$.
\end{itemize}
It can be presented in the form
\beq\label{xv}
\begin{array}{rclrcl}
A_\theta &=& \e^{-\frac{V}{2}}\pdf_\theta \e^{\frac{V}{2}} \; ;\qquad &
A_{\bar\theta} &=& \e^{\frac{V}{2}}\pdf_{\bar\theta} \e^{-\frac{V}{2}} \; ;\\
A_m &=& \e^{\frac{V}{2}}\left(\phi_m + \nabla_m\right) \e^{-\frac{V}{2}} \;
                       ;\qquad &
A_{\bar m} &=& \e^{-\frac{V}{2}}\left(\phi_{\bar m} + \nabla_{\bar m}\right)
 \e^{\frac{V}{2}} \; ;\\
L_{mn} &=& \e^{-\frac{V}{2}} M_{mn} \e^{\frac{V}{2}} \; ;\qquad &
L_{\bar m\bar n} &=& \e^{\frac{V}{2}} M_{\bar m\bar n} \e^{-\frac{V}{2}} \; .\\
\end{array}
\eeq
The constraint superfields $V$, $\phi_m$, $\phi_{\bar m}$, $M_{mn}$, and
$M_{\bar m\bar n}$ are determined up to local chiral transformations.

\subsec{BRS Symmetry in Superspace}
\label{brst}
It is convenient to describe the chiral transformations by a BRS (nilpotent)
operation. For this purpose one introduces a pair of chiral\hy antichiral,
anticommuting superfields $C$, $C^\kr$ with $\pdf_{\bar\theta} C = \pdf_\theta
C^\kr = 0$. The unconstrained prepotentials transform as follows:
\beq\label{xvi}
\begin{array}{rclrcl}
\s \e^V &=& \e^V C + C^\kr \e^V \; ;\\
\s \phi_m &=& {\cal D}_m C \; ;\qquad &
\s \phi_{\bar m} &=& -{\cal D}_{\bar m} C^\kr \; ;\\
\s M_{mn} &=& \{ C^\kr, M_{mn} \} \; ;\qquad &
\s M_{\bar m\bar n} &=& -\{ C, M_{\bar m\bar n} \} \;\\
\end{array}
\eeq
where ${\cal D}_m$ (or ${\cal D}_{\bar m}$ resp.) is the gauge covariant
derivative constructed with $\phi_m$ (or $\phi_{\bar m}$ resp.). The gauge
symmetry is represented by an anti\hy Hermitean Faddeev\Et Popov ghost
superfield
\beq\label{xvii}
K = \frac{C - C^\kr}{2} + \tanh\pounds_{\frac{V}{4}}\left(
\frac{C + C^\kr}{2} \right) \;
\eeq
where $\pounds_X = [X,\;]$ denotes the Lie bracket of the superfield $X$. The
`matter' transforms as
\beq\label{xviii}
\begin{array}{c}
\s \Lambda = - [ K, \Lambda ] \; ;\\
\s X_{mn} = - \{ K, X_{mn} \} \; ;\qquad
\s X_{\bar m\bar n} = - \{ K, X_{\bar m\bar n}\} \; .
\end{array}
\eeq
The local chiral symmetry is fixed if only the corresponding connections
$\phi_m$, $\phi_{\bar m}$ obey subsidiary conditions, \eg\ $\nabla^m\phi_m =
\nabla_m\phi^m = 0$. In order to find the gauge fixing and Faddeev\Et Popov\hy
terms one introduces a pair of chiral\hy antichiral (commuting) superfields
$D$, $D^\kr$ with $\pdf_{\bar\theta} D = \pdf_\theta D^\kr = 0$, as well as
their BRS variations $B = \s D$ and $B^\kr = \s D^\kr$. The superfields $B$,
$B^\kr$ are anticommuting and form a chiral\hy antichiral pair. They serve as
Lagrange multipliers for the gauge fixing conditions characterized by the real
parameter $\alpha$. The BRS (trivial) term of the action is
\beq\label{xix}
\begin{array}{rcl}
{\cal S}' &=& -\quart\,\s\ds\int_{\cal K}\md^2 z \md^2{\bar
z}\,g\,\Bigl(\pdf_\theta\Tr\left\{D \left(\nabla^m \phi_m
             - \alpha\pdf_\theta B^{\mbox{\rm \vphantom{\kr}}}
\right)\right\}\Bigr.\\
&& + \Bigl.\pdf_{\bar\theta}\Tr\left\{D^\kr \left(\nabla_m \phi^m
          - \alpha\pdf_\theta B^\kr \right)\right\}\Bigr) \; .
\end{array}
\eeq
The total action is $\widetilde S + S'$ and represents the starting point of
all perturbative or non\hy perturbative considerations.

\subsec{Correlation Functions}
\label{correl}
We would like to study correlation functions of the form
\beq\label{xx}
\vev{\prod_i {\cal O}_i} = \int [\md\widetilde{\mu}]\,\prod_i {\cal O}_i\,
              \exp\left\{-\frac{1}{e^2}\left(\widetilde S + S'\right)
                                     \right\}\;
\eeq
where ${\cal O}_i$ are gauge invariant ($\s {\cal O}_i = 0$) and metric
independent polynomials in the fields, $[\md\widetilde{\mu}]$ denotes the path 
integral
measure of all unconstrained superfields, \ie\ $V$, $\phi_m$, $\phi_{\bar m}$,
$M_{mn}$, $M_{\bar m\bar n}$, $X_{mn}$, $X_{\bar m\bar n}$, $\Lambda$, $C$,
$C^\kr$, $D$, $D^\kr$, $B$, and $B^\kr$; $e$ is the gauge coupling constant.

If one assumes that $\q {\cal O}_i = \qq {\cal O}_i = 0$, the correlation
functions are independent of the coupling constant. The reason for this
property is the form of the total action
\beq\label{xxi}
\widetilde S + S' = \q\qq{\cal V} + \q{\cal W} - \qq\bar{\cal W}
\eeq
where ${\cal V}$ is a Hermitean superfield and ${\cal W}$, $\bar{\cal W}$ is a
pair of complex conjugate chiral\hy antichiral anticommuting superfields, \ie\
$\qq{\cal W} = \q\bar{\cal W} = 0$.

Many observables are constructed from the (total) action by operations which
commute with both $\q$ and $\qq$ and therefore have the form~(\ref{xxi}). They
are highest components of BRS invariant superfields. Of course, they are $\q$
and $\qq$ invariant. Observables which are highest components can be shown 
to have vanishing correlation functions only.

\sect{Symmetries}
\label{symme}
Most of the symmetries of supersymmetric field theories are encoded in the
supercurrent \cite{Ferr}, a multiplet containing the energy\hy momentum tensor,
the
supersymmetry and the $\bbbr$\hy symmetry currents. For $N=2$ supersymmetry
an additional isovector current \cite{Sohn} corresponding to the automorphism
$SU(2)$
symmetry group belongs to the supermultiplet. (Of course, the number of
supersymmetry currents is doubled.)

The twisting procedure enables one to get the above currents for the TYM and
to recast the result into appropriate superfields \cite{Dah}, \cite{Marc}. The
method has
the advantage to be easily applicable \cite{Alva} to any $N=2$ supersymmetric
gauge
theory, \eg\ to super\hy Yang\Et Mills coupled to relaxed hypermultiplet
\cite{Howe}.

The approach we shall use below is adapted to the
relative simple structure of the TYM obtained from pure $N=2$ supersymmetric
Yang\Et Mills theory.

\subsec{Energy\hy Momentum Tensor}
\label{emt}
Consider a variation of the metric in the action~(\ref{xii}). The canonical
energy\hy momentum tensor defined by
\beq\label{iii_i}
\delta_g {\cal S} = - \eighth\int_{\cal K}\md^2z\md^2\bar z\,g\,
              \delta g^{\bar nm} \vartheta_{m\bar n}
\eeq
has the form
\beq\label{iii_ii}
\begin{array}{rcl}
\vartheta_{m\bar n} &=& \Tr\{-\imu k f_{m\bar n}
   + \D_m\lambda \D_{\bar n}\varphi + \D_m\varphi \D_{\bar n}\lambda
   -\imu\left(\psi_m \D_{\bar n} g_- + \psi_{\bar n} \D_m g_+\right)\\
&& -2\imu\lambda\{\psi_m,\psi_{\bar n}\} - g_{m\bar n}[k^2-\imu kf
   + \D_p\lambda \D^p\varphi + \D_p\varphi \D^p\lambda\\
&& -\imu\left(\psi_p \D^p g_- + \psi^p \D_p g_+\right)
   -2\imu\lambda\{\psi_p,\psi^p\} - \imu\varphi\{g_+,g_-\}\\
&& + [\varphi,\lambda]^2]\} \; .
\end{array}
\eeq
It is the last component of the superfield
$-2\imu\Tr\left\{\Lambda\left[F_{m\bar n} - g_{m\bar n}\left(F +
\imu[\D_\theta, \D_{\bar\theta}]\Lambda\right)\right]\right\}$. The on\hy
shell version of the latter reads
\beq\label{iii_iii}
Q_{m\bar n} = -2\imu\Tr\left\{\Lambda\left(F_{m\bar n} - \half g_{m\bar n} F
                   \right)\right\}
\eeq
and obeys the conservation laws
\beq\label{iii_iv}
\begin{array}{rcl}
\pdf_\theta\nabla^{\bar n} Q_{m\bar n} + \nabla^n J_{mn} &=& 0\; ;\\
\pdf_{\bar\theta}\nabla^n Q_{n\bar m} + \nabla^{\bar n} J_{\bar m\bar n}
   &=& 0
\end{array}
\eeq
where
\beq\label{iii_v}
\begin{array}{rcl}
J_{mn} &=& \imu\Tr\left\{\half F_{(m}{}^p X_{n)p} + 2 F_{m\theta} \D_n \Lambda
   - \Lambda \D_{[m} F_{n]\theta}\right\}\; ;\\
J_{\bar m\bar n} &=& \imu\Tr\left\{\half F^{\bar p}{}_{(\bar m}
   X_{\bar n)\bar p} - 2 F_{\bar m\bar\theta} \D_{\bar n} \Lambda
   + \Lambda \D_{[\bar m} F_{\bar n]\bar\theta}\right\}\; .\\
\end{array}
\eeq
In deriving these expressions we used the equations of motion~(\ref{xi});
hence $J_{mn}$ and its complex conjugate $J_{\bar m\bar n}$ have no definite
off\hy shell continuation like $Q_{m\bar n}$.

The energy\hy momentum tensor $\vartheta_{m\bar n}$ obeys the conservation
laws
\beq\label{iii_vi}
\nabla^n \vartheta_{mn} + \nabla^{\bar n} \vartheta_{m\bar n} = 0 \; ;\qquad
\nabla^n \vartheta_{n\bar m} + \nabla^{\bar n} \vartheta_{\bar m\bar n}
   = 0
\eeq
where
\beq\label{iii_vii}
\begin{array}{rcl}
\vartheta_{mn} &=& \Tr\left\{-\half f_{(m}{}^p f_{n)p}
   + \D_{(m} \lambda \D_{n)} \varphi + \imu\chi_{np} \D_m \psi^p
   - \imu\psi_n \D_m g_-\right\}\; ;\\
\vartheta_{\bar m\bar n} &=& \Tr\left\{\half f^{\bar p}{}_{(\bar m}
   f_{\bar n)\bar p} + \D_{(\bar m} \lambda \D_{\bar n)} \varphi
   + \imu\chi_{\bar n\bar p} \D_{\bar m} \psi^{\bar p}
   - \imu\psi_{\bar n} \D_{\bar m} g_+\right\}\; .\\
\end{array}
\eeq
The conservation partners $\vartheta_{mn}$, $\vartheta_{\bar m\bar n}$ of the
energy\hy momentum tensor are higher components of the superfields $J_{mn}$
and $J_{\bar m\bar n}$ but obviously not highest component like
$\vartheta_{m\bar n}$.

The correlation functions of $\vartheta_{m\bar n}$ are vanishing. From the
special structure of~(\ref{iii_iv}) it follows that the correlation functions
of all the components of $\nabla^n J_{mn}$, $\nabla^{\bar n} J_{\bar m\bar n}$
(and in particular of $\nabla^n \vartheta_{mn}$, $\nabla^{\bar n}
\vartheta_{\bar m\bar n}$) vanish.

Of course, radiative corrections may alter some of the above conclusions.
For Riemannian manifolds one can however show that the energy\hy momentum
tensor remains highest component. In particular, there is no contribution to
the energy\hy momentum trace coming from the (Riemannian) manifold
\cite{Dah_Lett}. We expect a similar property for TYM on K\"{a}hler 
manifolds.

\subsec{Fermionic Symmetries}
\label{fermi}
Consider the fermionic transformations~(\ref{ix}) with $z,\bar z$\hy
dependent parameters $\zeta$, $\xi$ and pick up terms proportional to
(K\"ahler) derivatives of these parameters. The fermionic symmetry currents
defined by
\beq\label{iii_ii_v}
\begin{array}{rcl}
\delta {\cal S} &=& \isixteenth\ds\int_{\cal K}\md^2 z\md^2\bar z\,g\,
   \Bigl[\nabla^m(\zeta + \imu\xi)\,s^{-\can}_m
       + \nabla^{\bar m}(\zeta + \imu\xi)\,s^{-\can}_{\bar m}\Bigr. \\
&& + \Bigl.\nabla^m(\zeta - \imu\xi)\,s^{+\can}_m
       + \nabla^{\bar m}(\zeta - \imu\xi)\,s^{+\can}_{\bar m}\Bigr]\\
\end{array}
\eeq
are
\beq\label{iii_ii_vi}
\begin{array}{rcl}
s^{+\can}_m &=& 2\Tr\left\{f_{mn} \psi^n - g_- \D_m \varphi\right\} \; ;\\
s^{+\can}_{\bar m} &=& 2\Tr\left\{\chi_{\bar m\bar n} \D^{\bar n} \varphi
   + \imu k\psi_{\bar m} + \varphi[\lambda,\psi_{\bar m}]\right\} \; ;\\
s^{-\can}_m &=& 2\Tr\left\{\chi_{mn} \D^n \varphi
   - \imu k\psi_m + \varphi[\lambda,\psi_m]\right\} \; ;\\
s^{-\can}_{\bar m} &=& 2\Tr\left\{f_{\bar m\bar n} \psi^{\bar n}
   - g_+ \D_{\bar m} \varphi\right\} \; .\\
\end{array}
\eeq
Due to the equations of motion~(\ref{xi}) they obey the conservation
rules
\beq\label{iii_ii_vii}
\nabla^m s^{\pm\can}_m + \nabla^{\bar m} s^{\pm\can}_{\bar m} = 0\; .
\eeq
In order to find the superfields whose components are related to fermionic 
symmetry
currents we investigate at first the global symmetries of the action.

\subsec{Global Symmetries}
\label{global}
The action~(\ref{x}) is invariant under a global $SU(2)$
\beq\label{iii_iii_viii}
\begin{array}{rclrcl}
\delta \psi_m &=& -v\psi_m - u_{mn}\psi^n\; ;\qquad &
\delta \psi_{\bar m} &=& v\psi_{\bar m} - u_{\bar m\bar n}\psi^{\bar n}\; ;\\
\delta \chi_{mn} &=& -v\chi_{mn} - u_{mn}g_-\; ;\qquad &
\delta \chi_{\bar m\bar n} &=&-v\chi_{\bar m\bar n}-u_{\bar m\bar n}g_+\; ;\\
\delta g_+ &=& -vg_+ + \half u_{mn}\chi^{mn}\; ;\qquad &
\delta g_- &=& vg_- + \half u_{\bar m\bar n}\chi^{\bar m\bar n}\;
\end{array}
\eeq
and under a global Abelian group:
\beq\label{iii_iii_ix}
\begin{array}{rclrcl}
\delta \psi_m &=& u \psi_m\; ;\qquad &
\delta \psi_{\bar m} &=& u \psi_{\bar m}\; ;\\
\delta \varphi &=& 2u\varphi\; ;\qquad &
\delta \lambda &=& -2u\lambda\; ;\\
\delta \chi_{mn} &=& -u\chi_{mn}\; ;\qquad &
\delta \chi_{\bar m\bar n} &=& -u\chi_{\bar m\bar n}\; ;\\
\delta g_+ &=& -u g_+\; ;\qquad &
\delta g_- &=& -u g_-\; .
\end{array}
\eeq
The parameters $u$, $v$ are real and $u_{mn}$, $u_{\bar m\bar n}$ are
antisymmetric and complex conjugate to each other, \ie\ $u_{\bar m\bar n} = -
\overline{u_{mn}}$. The corresponding currents are given by
\beq\label{iii_iii_x}
\begin{array}{rcl}
\delta {\cal S} &=& \eighth\ds\int_{\cal K}\md^2 z\md^2\bar z\,g\,
   \Bigl[\eighth\left(\nabla^p u^{mn} b_{mnp} + \nabla^{\bar p} u^{mn}
   b_{mn\bar p}\right.\Bigr.\\
&& \left. + \nabla^p u^{\bar m\bar n} b_{\bar m\bar np} + \nabla^{\bar p}
   u^{\bar m\bar n} b_{\bar m\bar n\bar p}\right)\\
&& \Bigl. -\imu \left(\nabla^m u b_m + \nabla^{\bar m} u b_{\bar m}\right)
   - \left(\nabla^m v b'_m + \nabla^{\bar m} v b'_{\bar m}\right)\Bigr]\; .
\end{array}
\eeq
The conserved combinations $b_m \pm b'_m$ and $b_{\bar m} \pm b'_{\bar m}$ are
responsible for two global Abelian symmetry groups $\bbbr_\pm$. Through their
charges they associate to each field two additive (real) quantum numbers
$r_\pm$.

One can still modify the above combinations of currents without disturbing
their conservation. A convenient choice is to improve $b_{\bar m} + b'_{\bar
m}$ to a $\q$ variation and $b_m - b'_m$ to a $\qq$ variation. The improved
currents, denoted by $b^+_{\bar m}$ and $b^-_m$ are defined by
\beq\label{iii_iii_xi}
b^+_{\bar m} = -4\imu\q\Tr\{\lambda\psi_{\bar m}\}\; ;\qquad
b^-_m = -4\imu\qq\Tr\{\lambda\psi_m\}\; .
\eeq
The last step is to construct the conservation partners $b^+_m$ and $b^-_{\bar
m}$. The superfields are then easily guessed to be
\beq\label{iii_iii_xii}
\begin{array}{rcl}
B^+_m &=& 4\imu\Tr\left\{\half F^n{}_{\bar\theta}
   X_{mn} - F_{\theta\bar\theta} \D_m\Lambda\right\}\; ;\\
B^+_{\bar m} &=& 4\imu\pdf_\theta\Tr\{\Lambda F_{\bar m\bar\theta}\}
\end{array}
\eeq
and
\beq\label{iii_iii_xiii}
\begin{array}{rcl}
B^-_m &=& 4\imu\pdf_{\bar\theta}\Tr\{\Lambda F_{m\theta}\}\; ;\\
B^-_{\bar m} &=& 4\imu\Tr\left\{\half F^{\bar n}{}_\theta
   X_{\bar m\bar n} - F_{\theta\bar\theta} \D_{\bar m}\Lambda\right\}\; .
\end{array}
\eeq
They obey the conservation rule
\beq\label{iii_iii_xiv}
\nabla^m B^\pm_m + \nabla^{\bar m} B^\pm_{\bar m} = 0\; .
\eeq
The first components of $B^\pm_m$, $B^\pm_{\bar m}$ are the $\bbbr_\pm$\hy
currents discussed above; none of them is observable since none of them is
simultaneously annihilated by both $\q$ and $\qq$. The next components are the
improved currents of the fermionic symmetries:
\beq\label{iii_iii_xv}
\begin{array}{rcl}
s^+_m &=& 2\Tr\left\{f_{mn}\psi^n + \varphi \D_m g_-\right\}\; ;\\
s^+_{\bar m} &=& -\Tr\left\{f \psi_{\bar m} + 2g_- \D_{\bar m} \varphi +
   2\varphi[\lambda,\psi_{\bar m}]\right\}\; ;\\
s^-_m &=& -\Tr\left\{-f \psi_m + 2g_+ \D_m \varphi + 2\varphi[\lambda,
   \psi_m]\right\}\; ;\\
s^-_{\bar m} &=& 2\Tr\left\{f_{\bar m\bar n}\psi^{\bar n} + \varphi \D_{\bar m}
   g_+\right\}\; .
\end{array}
\eeq
One can easily check that eqs.~(\ref{iii_iii_xv}) differ
from~(\ref{iii_ii_vi}) by terms which do not violate the conservation law.

Of the list~(\ref{iii_iii_xv}) only $s^+_{\bar m}$ and $s^-_m$ are
local observables since they are both $\q$ and $\qq$ invariant. However, being
the highest component of the superfields $4\imu\Tr\{\Lambda F_{\bar m\bar
\theta}\}$ and $4\imu\Tr\{\Lambda F_{m\theta}\}$ respectively, all their
correlation functions vanish.

There are two more conserved currents which correspond to commuting parameters
$u_{mn}$, $u_{\bar m\bar n}$ \cite{Dah}, \cite{Dem}. They give rise to the
antichiral
superfields
\beq\label{iii_iii_xvi}
\begin{array}{rcl}
B_{mnp} &=& 4\imu\Tr\left\{X_{mn} F_{p\theta}\right\}\; ;\\
B_{mn\bar p} &=& 8\imu g_{[m\bar p} \pdf_\theta \Tr\left\{\Lambda F_{n]\theta}
            \right\}
\end{array}
\eeq
and to their complex conjugates (chiral ones)
\beq\label{iii_iii_xvii}
\begin{array}{rcl}
B_{\bar m\bar np} &=& 8\imu g_{p[\bar m} \pdf_{\bar\theta} \Tr\left\{\Lambda
   F_{\bar n]\bar\theta}\right\}\; ;\\
B_{\bar m\bar n\bar p} &=& 4\imu\Tr\left\{X_{\bar m\bar n} F_{\bar
   p\bar\theta}\right\}\; .
\end{array}
\eeq
The conservation rules read
\beq\label{iii_iii_xviii}
\begin{array}{rcl}
\nabla^p B_{mnp} + \nabla^{\bar p} B_{mn\bar p} &=& 0\; ;\\
\nabla^p B_{\bar m\bar np} + \nabla^{\bar p} B_{\bar m\bar n\bar p} &=& 0\; .
\end{array}
\eeq

Two new local observables emerge:
\beq\label{iii_iii_xix}
\begin{array}{rcccl}
s_{mnp} &=& \lwc{\pdf_{\bar\theta} B_{mnp}} &=& 4\Tr\left\{\chi_{mn} \D_p
\varphi
   - f_{mn} \psi_p \right\}\; ;\\
s_{\bar m\bar n\bar p} &=& \lwc{\pdf_\theta B_{\bar m\bar n\bar p}} &=&
   4\Tr\left\{\chi_{\bar m\bar n} \D_{\bar p} \varphi - f_{\bar m\bar n}
\psi_{\bar
   p} \right\}\; .
\end{array}
\eeq
They are highest components of an antichiral or chiral superfield, 
respectively,
and therefore have vanishing correlation functions.

\subsec{Taking BRS into Account}
\label{BRS}
Also if BRS symmetry is taken into account, the trivial observables discussed
previously remain higher components of gauge invariant superfields. The
starting point is the total action $\widetilde {\cal S}+{\cal S}'$, which
includes
also the Lagrange multipliers $L_{mn}$, $L_{\bar m\bar n}$, the Faddeev\Et
Popov ghosts $C$, $C^\kr$, and their antighosts $D$, $D^\kr$.  
The dynamics of the total action is somewhat intricate---the superfields
$X_{mn}$, $X_{\bar m\bar n}$ are set equal to the corresponding Lagrange
multipliers
\beq\label{iii_iv_xxi}
X_{mn} = L_{mn}\; ;\qquad X_{\bar m\bar n} = L_{\bar m\bar n}\; ,
\eeq
therefore becoming covariantly chiral.
Some equations
of motion (see~(\ref{xi})) have to be modified:
\beq\label{iii_iv_xx}
\begin{array}{rcl}
\D^n L_{mn} + 2\D_\theta \D_m \Lambda + 2\imu\e^{-\frac{V}{2}}
   \left(\nabla_m B^\kr - \left[C^\kr,\nabla_m D^\kr\right]\right)
   \e^{\frac{V}{2}} &=& 0\; ;\\
\D^{\bar n} L_{\bar m\bar n} + 2\D_{\bar\theta} \D_{\bar m} \Lambda
   + 2\imu\e^{\frac{V}{2}}
   \left(\nabla_{\bar m} B + \left[C,\nabla_{\bar m} D\right]\right)
   \e^{-\frac{V}{2}} &=& 0\; .
\end{array}
\eeq

The gauge is fixed by
\beq\label{iii_iv_xxii}
\nabla^m \phi_m = 2\alpha\pdf_\theta B\; ;\qquad \nabla^{\bar m} \phi_{\bar m}
=
   2\alpha\pdf_{\bar\theta} B^\kr\; .
\eeq
Finally, ghosts and antighosts obey the equations
\beq\label{iii_iv_xxiii}
\begin{array}{rclrcl}
\nabla^m {\cal D}_m C &=& 0\; ;\qquad &
\nabla^{\bar m} {\cal D}_{\bar m} C^\kr &=& 0\; ;\\
{\cal D}_m \nabla^m D &=& 0\; ;\qquad &
{\cal D}_{\bar m} \nabla^{\bar m} D^\kr &=& 0\; .
\end{array}
\eeq

After some calculation one gets the following system of BRS modified current
superfields corresponding to the fermionic symmetries:
\beq\label{iii_iv_xxiv}
\begin{array}{rcl}
\widetilde B^+_m &=& 4\imu\Tr\left\{\half F^n{}_{\bar\theta}
   X_{mn} - F_{\theta\bar\theta} \D_m \Lambda + \imu\s\left(
   \nabla_m D^\kr\e^V\pdf_{\bar\theta}\e^{-V} \right) \right\} \; ;\\
\widetilde B^+_{\bar m} &=& 4\imu\Tr\left\{\pdf_\theta\left(\Lambda
   F_{\bar m\bar\theta}\right) - \imu\s\left(D^\kr \pdf_{\bar\theta}
   \phi_{\bar m}\right) \right\} \; ;  \\
\widetilde B^-_m &=& 4\imu\Tr\left\{\pdf_{\bar\theta}\left(\Lambda
   F_{m\theta}\right) - \imu\s\left(D \pdf_\theta \phi_m\right)
   \right\} \; ;  \\
\widetilde B^-_{\bar m} &=& 4\imu\Tr\left\{\half F^{\bar n}{}_\theta
   X_{\bar m\bar n} - F_{\theta\bar\theta} \D_{\bar m} \Lambda + \imu\s\left(
   \nabla_{\bar m} D\e^{-V}\pdf_\theta\e^V \right) \right\} \;.
\end{array}
\eeq
The observable currents are $\widetilde s^+_{\bar m}$ and $\widetilde s^-_m$.
The differences $\widetilde s^+_{\bar m} - s^+_{\bar m}$ and $\widetilde s^-_m
- s^-_m$ are gauge variations of the second component of the chiral
superfields $-4\Tr\{D^\kr
\pdf_{\bar\theta} \phi_{\bar m}\}$ and $-4\Tr\{D \pdf_\theta \phi_m\}$, 
respectively.

\sect{Perturbative Check}
\label{pertu}
One would like to have some confirmation about the correctness of the field
theoretic description of TYM on K\"{a}hler manifolds. In the present section
we shall compute in perturbation theory the gravitational contribution to
$\bbbr_{\pm}$\hy anomalies. The latter can be considered as finite radiative
corrections
to the conservation law for the Abelian currents. It will turn out that the
anomalies are equal.
When integrated over the K\"{a}hler manifold they both yield the (formal)
dimension
of the instanton moduli space. The resulting number is well known in the
mathematical literature and has a standard representation by local
polynomials in the curvature.

On the other hand $\bbbr_{\pm}$\hy anomalies get perturbative contributions
from all the (super)fields of the model. Hence, one can in fact check the
proposed superspace description.

Since our calculation is limited to one\hy loop approximation, we shall
regularize the $\bbbr_{\pm}$\hy currents by point\hy splitting. This way all
the symmetries which are supposed to be preserved at the quantum level will be
automatically taken into account.

\subsec{Green Functions}
\label{green}
The Green functions can be obtained from the linearized equations of motion
with sources. The source term that is added to the total action has the form
\beq\label{iv_i}
\begin{array}{c}
\quart\ds\int_{\cal K} \md^2 z \md^2{\bar z}g \left\{
{\pdf}_{\theta}
\Tr \left( \half K_{mn}M^{mn} + J^m{\phi}_m + J_{B}B + J_{C}C + J_{D}D
\right) \right. \; \\
 + {\pdf}_{\bar{\theta}} \Tr \left(
\half K^{mn}M_{mn} + J_m M^m + J_{B^\kr} B^\kr + J_{C^\kr} + J_{D^\kr}
D^\kr \right) \; \\
 + {\pdf}_{\theta}
{\pdf}_{\bar{\theta}} \Tr \left. \left( \half J^{mn}X_{mn} +
\half J_{mn}X^{mn} + J_{\Lambda}{\Lambda} + J V \right) \right\}
\end{array}
\eeq
where the sources $K_{mn}$, $K_{\bar m \bar n}$, $J_m$, $J_{\bar m}$, $J_B$,
$J_{B^\kr}$, $J_C$, $J_{C^\kr}$, $J_D$, $J_{D^\kr}$, $J_{mn}$,
$J_{\bar m \bar n}$, $J_{\Lambda}$, $J$ and the corresponding superfields
$M_{\bar m \bar n}$, $M_{mn}$, ${\phi}_{\bar m}$, ${\phi}_m$, $B$, $B^\kr$,
$C$, $C^\kr$, $D$, $D^\kr$, $X_{\bar m \bar n}$, $X_{mn}$, $\Lambda$, $V$ have
the same symmetry and reality (or chirality) properties.

The linearized equations of motion are
\[
\begin{array}{rcl}
\half{\nabla}^n M_{mn} + {\nabla}_m  \left(  B^\kr - \imu
{\pdf}_{\theta}\Lambda \right)         & = & J_m \; ; \\
\half{\nabla}^{\bar n} M_{\bar m \bar n} + {\nabla}_{\bar m}  \left( B -
\imu {\pdf}_{\bar{\theta}}\Lambda \right) & = & J_{\bar m} \; ;
\end{array}
\]
\beq\label{iv_ii}
\begin{array}{rclrcl}
\imu {\nabla}_{[ m}{\phi}_{n ]} + {\pdf}_{\bar{\theta}}X_{mn} &=& 2
K_{mn} \; ;\; &
\imu {\nabla}_{[ \bar m}
{\phi}_{\bar n ]} + {\pdf}_{\theta}X_{\bar m \bar n} &=& 2 K_{\bar m \bar n}
\; ; \\
X_{mn} &=& M_{mn} + 2 J_{mn} \; ;\; &
X_{\bar m \bar n} &=& M_{\bar m \bar n} + 2 J_{\bar m \bar n} \; ; \\
{\nabla}^m{\phi}_m  &=& 2 \alpha {\pdf}_{\theta}B + J_{B} \; ; \; &
{\nabla}_m {\phi}^m
&=& 2 \alpha {\pdf}_{\bar{\theta}}B^\kr + J_{B^\kr} \; ;\\
{\nabla}_m {\nabla}^m C &=& J_{D} \; ;\; &
{\nabla}^m{\nabla}_m
C^\kr &=& J_{D^\kr} \; ;\\
{\nabla}_m{\nabla}^m D &=& J_{C} \; ;\; &
{\nabla}^m{\nabla}_m D^\kr &=& J_{C^\kr} \; ;
\end{array}
\eeq
\[
\begin{array}{rcl}
{\nabla}_m {\nabla}^m V  + {\nabla}_m {\phi}^m -
                       {\nabla}^m {\phi}_m + 4\imu {\pdf}_{\theta}
{\pdf}_{\bar{\theta}} \Lambda + \imu J_{\Lambda} &=& 0 \; ;\\
{\nabla}_m {\nabla}^m \Lambda + \imu J        & = & 0  \; .
\end{array}
\]
Eqs.~(\ref{iv_ii}) can be converted into differential equations for the Green
functions (GF). The solution can be expressed in terms of two basic GF, the
scalar GF $G(z, z')$ and the vector one $G_{m {\bar n}'}(z, z')$ the 
definition of which is given by 
\beq\label{iv_iii}
\begin{array}{rcl}
{\nabla}_{p}{\nabla}^{p}G(z, z') & = & - \delta (z, z') \; ; \\
{\nabla}_{p}{\nabla}^{p}G_{m {\bar n}'}(z, z') & = & - g_{m {\bar n}'}( z, z' )
\delta (z, z')
\end{array}
\eeq
where
\beq\label{iv_iv}
\delta (z, z') = {g}^{-1}{\delta}^2 (z - z'){\delta}^2({\bar z}-
{\bar z}') \; .
\eeq
The factor $g_{m {\bar n}'}( z, z' )$ is explained below in
section~\ref{short}.

By taking ${\cal K}$ a compact K\"ahler manifold and by observing that the
Laplacian ${\nabla}_{p}{\nabla}^{p}$ is an elliptic operator one can assume
that eqs.~(\ref{iv_iii}) have unique solutions.

We choose to work in the gauge $\alpha = 1$. Below we list only those Green
functions which are necessary for the computation of anomalies:
\begin{eqnarray}\label{iv_v}
\vev{X_{mna} {\phi}_{{\bar p}'b}}       & = & 2 \imu {\delta}_{ab} (
\bar{\theta} -
{\bar{\theta}}' ) {\nabla}_{[ m} G_{n ] {\bar p}'}( z, z' ) \; ; \nonumber \\
\vev{X_{\bar m \bar n a} {\phi}_{p' b}} & = & 2 \imu {\delta}_{ab} ( \theta -
{\theta}' ) {\nabla}_{[ \bar m} G_{p' \bar n ]} ( z, z' ) \; ; \nonumber \\
\vev{{\Lambda}_a {V'}_b }               & = & \imu {\delta}_{ab} ( \theta -
{\theta}' ) ( \bar{\theta} - {\bar{\theta}}' ) G( z, z' ) \; ; \nonumber \\
\vev{{\phi}_{ma} {B'}_b }               & = & - {\delta}_{ab} ( \theta -
{\theta}' ) {\nabla}_m G( z, z' ) \; ; \nonumber \\
\vev{{\phi}_{\bar m a} {{\bar B}'}_b }  & = & - {\delta}_{ab} ( \bar{\theta} -
{\bar{\theta}}' ) {\nabla}_{\bar m} G( z, z' ) \; ; \\
\vev{V_a {B'}_b }                       & = & - {\delta}_{ab} ( \theta -
{\theta}' ) G( z, z' ) \; ; \nonumber \\
\vev{V_a {{\bar B}'}_b }                & = & - {\delta}_{ab} ( \bar{\theta} -
{\bar{\theta}}' ) G( z, z' ) \; ; \nonumber \\
\vev{C_a {D'}_b }                       & = & {\delta}_{ab} ( \theta -
{\theta}' ) G( z, z' ) \; ; \nonumber \\
\vev{{\bar C}_a {{\bar D}'}_b }         & = & {\delta}_{ab} ( \bar{\theta} -
{\bar{\theta}}' ) G( z, z' ) \; .\nonumber
\end{eqnarray}
The brackets on the left hand side of~(\ref{iv_vi}) indicate that the two\hy
point function has to be calculated with the formula
\beq\label{iv_vi}
\vev{\ldots } = \int [\md\widetilde{\mu}]\,\ldots\,\exp \left\{ -\frac{1}{e^2}
{\cal S}_{\mbox{\sevenrm lin}} \right\} \; .
\eeq
The linearized action ${\cal S}_{\mbox{\sevenrm lin}}$ leads to the
equations~(\ref{iv_ii}), albeit with the source superfields set equal to zero.

\subsec{Short Distance Behaviour}
\label{short}
GF have singularities in the scalar variable $\sigma$, \ie\ one fourth of
the geodetic interval squared between the points $( z^m , z^{\bar m} )$ and
$( z^{m'} , z^{{\bar m}'} )$. The variable satisfies
\beq\label{iv_vii}
\sigma = g^{\bar n m} {\sigma}_m {\sigma}_{\bar n} \;
\eeq
where
\beq\label{iv_viii}
{\sigma}_m = {\nabla}_m \sigma \; ; \qquad \qquad {\sigma}_{\bar m} =
{\nabla}_{\bar m} \sigma \; .
\eeq

The residue of the singularity involves the so called parallel
displacement matrix
$g_{m {\bar n}'}( z, z' )$ \cite{DWitt} defined by
\beq\label{iv_ix}
\left( {\sigma}^p {\nabla}_p + {\sigma}^{\bar p} {\nabla}_p \right)
g_{m {\bar n}'}( z, z' ) = 0 \; ; \qquad \left[ g_{m {\bar n}'} \right]
= g_{m \bar n} \; .
\eeq
The square bracket used in the boundary condition means coinciding
arguments, \ie\ $z^{m'} = z^m$ and $z^{{\bar m}'} = z^{\bar m}$.

The scalar Green function has the following short distance behaviour
\cite{DWitt}:
\beq\label{iv_x}
G( z, z') = - \frac{\Gamma ( z, z' )}{64{\pi}^2} \left\{
\frac{1}{\sigma} + [v_0]
\ln \sigma + {\cal O}( {\sigma}_m \ln \sigma ) \right\} \;
\eeq
valid up to terms of at least first order in ${\sigma}_m \ln \sigma$ or
${\sigma}_{\bar m} \ln \sigma$. The functions $\Gamma ( z, z' )$ and
$v_0 ( z, z' )$ can be determined from differential equations with boundary
conditions that are similar to~(\ref{iv_ix}). The only relation we shall need
in the following is:
\beq\label{iv_xi}
\left( {\nabla}_m + g_{m {\bar n}'} {\nabla}^{{\bar n}'} \right) \ln \Gamma
( z, z' ) = {\cal O}_2 ( {\sigma}_m , {\sigma}_{\bar m} ) \; ;
\quad [\Gamma] = 1 \;
\eeq
where the symbol ${\cal O}_2$ on the right hand side of the first
equation~(\ref{iv_xi}) means terms at least bilinear in ${\sigma}_m$ and
${\sigma}_{\bar m}$.

Instead of the vector GF it proves convenient to introduce \cite{Chri} two
bilocal
unprimed tensors
\beq\label{iv_xii}
\begin{array}{rcl}
{\overline{G}}_{m \bar n} ( z, z' ) & = & g^{{\bar p}'}_{\bar n} ( z, z' )
G_{m {\bar p}'} ( z, z' ) \; ; \\
{\widetilde{G}}_{m \bar n} ( z, z' ) & = & g^{p'}_m ( z, z' ) G_{{p}' \bar n}
( z, z' ) \; .
\end{array}
\eeq
The short distance expansion of the vector GF reads now
\beq\label{iv_xiii}
{\overline{G}}_{m \bar n} ( z, z' ) = g_{m \bar n} G ( z, z' ) -
\frac{\Gamma ( z, z' )}{128{\pi}^2} R_{m \bar n} \ln \sigma +
{\cal O} ({\sigma}_m \ln \sigma ) \;
\eeq
where $R_{m \bar n}$ is the Ricci tensor of the K\"{a}hler manifold.

\subsec{Regularization of Superfield Currents}
\label{regul}
We assume that the fermionic symmetries are preserved in perturbation theory.
The corresponding currents $s^{\pm}_m$, $s^{\pm}_{\bar m}$ receive radiative
corrections, can be, however, redefined as to remain conserved. 
As a consequence one
can use the superspace approach for quantum computations. However,
the conservation law~(\ref{iii_iii_xiv}) of the superfield currents
$B^{\pm}_m$, $B^{\pm}_{\bar m}$ is violated in perturbation theory. The
breakdown of current conservation gives rise to $\bbbr_{\pm}$\hy anomalies.

For evaluating the gravitational contribution it is sufficient to compute the
vacuum expectation value of the left hand side of the conservation
law~(\ref{iii_iii_xiv}). Here we consider explicitly only the $\bbbr_{+}$\hy
anomaly. The appropriate superfield currents were given
in~(\ref{iii_iv_xxiv}).  In the one\hy loop approximation one uses the
linearized expressions
\beq\label{iv_xiv}
\begin{array}{rcl}
{\widetilde{B}}^{+}_{m}      & = & 4\imu\Tr\Bigl\{ \frac{1}{2}g^{\bar p n}
X_{mn}{\pdf}_{\bar{\theta}}\left( {\nabla}_{\bar p}V + {\phi}_{\bar n}
\right)  + {\pdf}_{\theta}{\pdf}_{\bar{\theta}}V{\nabla}_m \Lambda
\Bigr. \; \\
      &   & + \imu\Bigl. \left( {\nabla}_m D^\kr
{\pdf}_{\bar{\theta}}C^\kr - {\nabla}_m B^\kr{\pdf}_{\bar{\theta}}V
\right) \Bigr\} \; ;
\end{array}
\eeq
\beq\label{iv_xv}
\begin{array}{rcl}
{\widetilde{B}}^{+}_{\bar m} & = & 4 \imu\Tr\Bigl\{ {\pdf}_{\bar{\theta}}
\left( {\nabla}_{\bar m}V + {\phi}_{\bar m} \right) {\pdf}_{\theta}\Lambda
 - \Lambda{\pdf}_{\theta}{\pdf}_{\bar{\theta}}{\nabla}_{\bar m}V
\Bigr. \; \\
                               &   & - \imu\Bigl. \left( D^\kr
{\pdf}_{\bar{\theta}}{\nabla}_{\bar m}C^\kr + B^\kr
{\pdf}_{\bar{\theta}}{\phi}_{\bar m} \right) \Bigr\} \; .
\end{array}
\eeq

The linearized currents are regularized by symmetric point\hy splitting being 
summarized in the following rules:
\begin{enumerate}
\item
For each term of~(\ref{iv_xiv}) and~(\ref{iv_xv}) one takes into account both
factor orderings with equal weight.
\item
Each primed index, \ie\ corresponding to the coordinates $( z^{m'},
z^{{\bar m}'} )$ is accompanied by a parallel displacement matrix
element $g_{m{\bar n}'}(z, z')$ or by its inverse $g^{{\bar n}'m}(z', z )$.
\end{enumerate}

As an example let us write down the regularized version of~(\ref{iv_xiv})
\beq\label{iv_xvi}
\begin{array}{rcl}
{\widetilde{B}}^{+}_m ( z, z' ) & = & 2 \imu\Tr\Bigl\{\half\left(
g^{{\bar p}' n}X_{mn}{\pdf}_{{\bar{\theta}}'}{\phi}_{{\bar p}'} -
g^{n'}_m g^{{\bar p} r'}
{\pdf}_{\bar{\theta}}{\phi}_{\bar p} X_{n' r'} \right) \Bigr. \;
\\
                                  &   & + {\nabla}_m \Lambda
{\pdf}_{{\theta}'} {\pdf}_{{\bar{\theta}}'} V' + g^{n'}_m
{\pdf}_{\theta} {\pdf}_{\bar{\theta}} V {\nabla}_{n'} {\Lambda}' \;
\\
                                  &   & + \imu\left({\nabla}_m D^\kr
{\pdf}_{{\bar{\theta}}'}C^{\prime\kr} + g^{n'}_m {\pdf}_{\bar{\theta}} C^\kr
{\nabla}_{n'} B^{\prime\kr} \right) \; \\
                                  &   & - \imu\Bigl.\left( {\nabla}_m
B^\kr {\pdf}_{{\bar{\theta}}'} V' - g^{n'}_m {\pdf}_{\bar{\theta}} V
{\nabla}_{n'} B^{\prime\kr} \right) \Bigr\} \; .
\end{array}
\eeq
One finds a similar expression for ${\widetilde{B}}^{+}_{\bar m} ( z, z' )$.
The vacuum expectation values of the regularized currents are
\beq\label{iv_xvii}
\begin{array}{rcl}
\vev{{\widetilde{B}}^{+}_m (z, z')}        & = & 2 n \Bigl\{ g^{{\bar p} n'}
{\nabla}_{[ m} G_{n ]{\bar p}'} - {\nabla}_m G \Bigr. \; \\
                                         &   & + \Bigl. g^{n'}_m \left(
g^{{\bar p} r'}{\nabla}_{[ n'}G_{r' ] \bar p} -  {\nabla}_{n'} G
\right) \Bigr\} \; ; \\
\vev{{\widetilde{B}}^{+}_{\bar m} (z, z')} & = & 0 \; .
\end{array}
\eeq

Recalling the
meaning of the square bracket in eqs.~(\ref{iv_ix}) and~(\ref{iv_x}) the 
gravitational contribution to the $\bbbr_{+}$\hy anomaly becomes
\beq\label{iv_xviii}
B^{(+)} (z) = \frac{1}{2} {\nabla}^m \left[ \vev{ {\widetilde{B}}^{+}_m (z, z')
} \right] \;
\eeq
can be evaluated with the help of Synge's theorem \cite{Chri}
\beq\label{iv_xix}
B^{(+)} (z) = \frac{1}{2} \left[ \vev{ \left( {\nabla}^m + g^{{\bar n}' m}
{\nabla}_{{\bar n}'} \right){\widetilde{B}}^{+}_m (z, z') } \right]
\; .
\eeq
If one inserts~(\ref{iv_xvii}) into~(\ref{iv_xix}) and if one tries to exhibit
the combinations~(\ref{iv_xii}) in place of the vector GF, one gets
\beq\label{iv_xx}
\begin{array}{rcl}
B^{(+)} (z) & = & n \left[ \left\{ \left( g^{\bar n p'} {\nabla}_{[ \bar s}
g_{p' \bar u ]} - g_{p' [ \bar s} g_{r' \bar u ]} {\nabla}^{p'} g^{\bar n r'}
\right) \left( {\nabla}^{\bar s}\, {\overline{G}}^{\bar u}_{\bar n} +
{\nabla}^{\bar s} {\widetilde{G}}^{\bar u}_{\bar n} \right. \right. \right. \;
\\
            &   & + \left. g^{{\bar t}'}_{\bar n} {\nabla}^{\bar s}
g^{\bar v}_{{\bar t}'} {\overline{G}}^{\bar u}_{\bar v} - g^{\bar s t'}
g^{\bar u v'} {\nabla}_{t'} g^w_{v'} {\widetilde{G}}_{w \bar n} \right) \;
\\
            &   & + \left( {\nabla}_{{\bar m}'} g^{{\bar m}'}_{\bar s}
- g_{n' \bar s} {\nabla}_{\bar m} g^{\bar m n'} \right) \left( 2
g^{\bar s u'} {\nabla}_{u'} G  + {\nabla}^{[ \bar s}
{\widetilde{G}}^{\bar u ]}_{ \bar u} \right. \; \\
            &   & - \left. \left. \left.  g^{[ \bar s t'}
g^{\bar u ] v'} {\nabla}_{t'} g^p_{v'} {\widetilde{G}}_{p \bar u}
\right) \right\} \right] \; .
\end{array}
\eeq

Now one uses the short distance expansions~(\ref{iv_x}) and~(\ref{iv_xiii})
in~(\ref{iv_xx}). One realizes immediately that only singularities of the form
$ {\sigma}^{-2}$ contribute to~(\ref{iv_xx}). The residues can be evaluated by
means of the formulae:
\beq\label{iv_xxi}
\begin{array}{l}
g^{p'}_n {\nabla}_{[ \bar s} g_{p' \bar u ]} -
g_{p' [ \bar s} g_{r' \bar u ]} {\nabla}^{p'} g^{r'}_n \\
\qquad\qquad = \ds\frac{1}{12}
{\sigma}^a {\sigma}^b {\sigma}^{\bar c} \left( {R^r}_{ab[ \bar s}
R_{\bar u ] rn \bar c} - {R^r}_{an [ \bar s} R_{\bar u ] rb \bar c} \right)
+ \ldots \\
{\nabla}_{{\bar m}'} g^{{\bar m}'}_{\bar s} -
g_{n' \bar s} {\nabla}_{\bar m} g^{\bar m n'} \\
\qquad\qquad =\ds - \frac{1}{6}
{\sigma}^a {\sigma}^b {\sigma}^{\bar c} \left( \frac{1}{2} {R^p}_{ra \bar s}
{R^r}_{pb \bar c} + \frac{1}{4} {R^p}_{ab \bar c} R_{p \bar s} + {R^p}_a
R_{\bar c pb \bar s} \right) + \ldots
\end{array}
\eeq
where the dots collect all the terms which are unimportant for the present
calculation. The result reads
\beq\label{iv_xxii}
\begin{array}{rcl}
B^{(+)} (z) & = &\ds \frac{n}{128{\pi}^2} \Biggl[
\frac{{\sigma}^a{\sigma}^{\bar b}}{\sigma}
\Biggl\{ \frac{{\sigma}^c{\sigma}^{\bar d}}{3 \sigma} \left(
{R^p}_{ac \bar b} R_{p \bar d} + {R^p}_a R_{\bar b pc \bar d} \right. \Biggr.
\Biggr.\\
            &   & - \Biggl. \Biggl. \left. {R^p}_{ra \bar b}
{R^r}_{pc \bar d}
- {{R^p}_{ac}}^r R_{\bar b pr \bar d} \right) + {R^p}_a R_{p \bar b} -
R R_{a \bar b}  \Biggr\} \Biggr] \; .
\end{array}
\eeq
The average over the directions of ${\sigma}_m$, ${\sigma}_{\bar m}$
amounts to the following replacements:
\beq\label{iv_xxiii}
{\sigma}^a{\sigma}^{\bar b} {\rightarrow} g^{\bar b a} \frac{\sigma}{2} \;
; \qquad
{\sigma}^a {\sigma}^b {\sigma}^{\bar c}{\sigma}^{\bar d} {\rightarrow} \left(
g^{\bar c a} g^{\bar d b} + g^{\bar d a} g^{\bar c b} \right)
\frac{{\sigma}^2}{6} \; .
\eeq
One can avoid the average if one performs the subtraction of certain
direction dependent terms \cite{Niel} in the regularized current.

As a result the gravitational contribution to the $\bbbr_{+}$\hy anomaly 
assumes 
the manifestly local form
\beq\label{iv_xxiv}
B^{(+)} = \frac{n}{64{\pi}^2} \left( \frac{1}{3} {R^a}_b {R^b}_a - \frac{1}{4}
R^2 - \frac{1}{12} {{R^a}_{bc}}^d {{R^b}_{ad}}^c \right) \; .
\eeq
It turns out that the contribution $B^{(-)}$ of the external gravity to the
$\bbbr_{-}$\hy anomaly has the same expression.

Before discussing the result~(\ref{iv_xxiv}) let us comment on its
derivation. The contribution of the Faddeev\Et Popov ghosts cancels
in~(\ref{iv_xxiv}), hence the same result is obtained if one keeps only the
first three terms in~(\ref{iv_xiv}) and~(\ref{iv_xv}). The cancellation is due
to the special interplay between BRS and fermionic symmetries.  
suggests that the calculation could be performed 
without Faddeev\Et Popov ghosts, but with the ghosts 
introduced in section~\ref{descent}.

Eq.~(\ref{iv_xxiv}) can be written in the more familiar form
\beq\label{iv_xxv}
B^{(\pm)} = n ( H - E ) \; .
\eeq
Here $E$ and $H$ are the K\"{a}hler analogs of the Euler and Hirzebruch
(signature) densities:
\beq\label{iv_xxvi}
\begin{array}{rcl}
E & = &\ds \frac{1}{128{\pi}^2} \left( {{R^m}_{pr}}^n {{R^p}_{mn}}^r -
2 {R^m}_n {R^n}_m + R^2 \right) \; ; \\
H & = &\ds \frac{1}{192{\pi}^2} \left( {{R^m}_{pr}}^n {{R^p}_{mn}}^r -
{R^m}_n {R^n}_m \right) \; .
\end{array}
\eeq
Recall that on a Riemannian manifold without boundary, the
densities~(\ref{iv_xxvi}) have the expressions \cite{Eguc}
\beq\label{iv_xxvii}
E = \frac{1}{32{\pi}^2} {\widetilde{R}}_{\lambda \mu \nu \rho}
{\widetilde{R}}^{\nu \rho \lambda \mu} \; ; \qquad
H = \frac{1}{48{\pi}^2} {\widetilde{R}}_{\lambda \mu \nu \rho}
R^{\nu \rho \lambda \mu} \;
\eeq
where $R_{\lambda \mu \nu \rho}$ is the curvature tensor and
${\widetilde{R}}_{\lambda \mu \nu \rho}$ its dual
\beq\label{iv_xxviii}
{\widetilde{R}}_{\lambda \mu \nu \rho} = \frac{1}{2}
\sqrt{g}\,{\epsilon}_{\lambda \mu \alpha \beta}
{{R}_{\nu \rho}}^{\alpha \beta} \; .
\eeq

One can write the curvature tensor $R_{\lambda \mu \nu \rho}$ in spinorial
form and decompose it into irreducible spinors \cite{Wess}, \cite{Wald}. In
passing to
K\"{a}hler four\hy manifolds one realizes that the curvature tensor has the
structure $R_{\bar m np \bar r}$, symmetric in $\bar m$, $\bar r$ and $n$,
$p$ respectively. In terms of irreducible spinors it has the form
\beq\label{iv_xxix}
\begin{array}{rcl}
R_{\bar m pr \bar n} & = & 4 \left\{ 2 \left( g_{p \bar m} g_{r \bar n} +
g_{r \bar m} g_{p \bar n} \right) U - e^{\alpha}_{\bar m} e^{\beta}_p
e^{\gamma}_r e^{\delta}_{\bar n} U_{\alpha \beta \gamma \delta}
\right. \; \\
                     &   & + \left. \left( e^{\alpha}_{\bar m}
e^{\beta}_p g_{r \bar n} + e^{\alpha}_{\bar n} e^{\beta}_r g_{p \bar m} \right)
U_{\alpha \beta} \right\} \; .
\end{array}
\eeq
Here $U$, $U_{\alpha \beta}$ and $U_{\alpha \beta \gamma \delta}$ are
irreducible spinors, completely symmetric in their indices. The zweibeins
$e^{\alpha}_m$, $e^{\beta}_{\bar n}$ convert spinor indices into holomorphic
and anti\hy holomorphic ones.

The densities~(\ref{iv_xxvi}) can also be written in terms of irreducible
spinors. For K\"{a}hler manifolds one finds:
\beq\label{iv_xxx}
\begin{array}{rcl}
E & = &\ds \frac{1}{8{\pi}^2} \left( U_{\alpha \beta \gamma \delta}
U^{\alpha \beta \gamma \delta} + 4 U_{\alpha \beta} U^{\alpha \beta}
+ 48 U^2 \right) \; ; \\
H & = &\ds \frac{1}{12{\pi}^2} \left( U_{\alpha \beta \gamma \delta}
U^{\alpha \beta \gamma \delta} - 24 U^2 \right) \; .
\end{array}
\eeq
By using~(\ref{iv_xxix}) it is possible to express~(\ref{iv_xxx}) through the
curvature tensor over the K\"{a}hler manifold. The result is~(\ref{iv_xxvi}).

\sect{Donaldson Cohomology}
\label{donal}
Let $({\cal K}, \gamma)$ be a compact K\"{a}hler four\hy manifold with
K\"{a}hler form $\gamma$ given by~(\ref{i}). Let $\E$ be a complex vector
bundle with structure group $\G$ assumed Lie, compact and
semi\hy simple. The connection on $\E$ splits into a $(1,0)$ part $a =
a_{m}\md z^m$ and a $(0,1)$ part ${\bar a} = a_{\bar m}\md z^{\bar m}$. The
curvature two\hy form can be decomposed into its $(2,0)$, $(0,2)$ and $(1,1)$
parts as follows:
\beq\label{v_i}
\begin{array}{c}
f^{(2,0)} = {\pdf}a + a^2 \; ; \qquad f^{(0,2)} =
{\bar{\pdf}}{\bar a} + {\bar a}^2 \; ; \\
f^{(1,1)} = {\pdf}{\bar a} + {\bar{\pdf}}a + \{ a, \bar a
\} \; .
\end{array}
\eeq
They obey the Bianchi identities
\beq\label{v_ii}
\begin{array}{c}
\D f^{(1,1)} + {\Dq} f^{(2,0)} = 0 \; ; \qquad {\Dq} f^{(1,1)} +
\D f^{(0,2)} = 0 \; ; \\
\D f^{(2,0)} = \Dq f^{(0,2)} = 0 \; .
\end{array}
\eeq
By using~(\ref{v_ii}) one derives the basic identities (For simplicity we limit 
ourselves to invariant polynomials quadratic in the curvature)
\beq\label{v_iii}
\begin{array}{rcl}
{\pdf}\; \Tr \left(\half {f^{(1,1)}}^2 + f^{(2,0)}
f^{(0,2)} \right) + {\bar{\pdf}}\; \Tr f^{(1,1)} f^{(2,0)} & = & 0
\; ; \\
{\bar{\pdf}}\; \Tr \left(\half {f^{(1,1)}}^2 + f^{(2,0)}
f^{(0,2)} \right) + {\pdf}\; \Tr f^{(1,1)} f^{(0,2)} & = & 0 \; .
\end{array}
\eeq
Obviously, the last terms in both eqs.~(\ref{v_iii}) vanish, rendering the
invariant $(2,2)$\hy form $\Tr \left( \frac{1}{2} f^{(1,1) \; 2} +
f^{(2,0)}f^{(0,2)} \right)$ closed with respect to both $\pdf$ and
$\bar{\pdf}$.

Locally, the closed form can be represented as
\beq\label{v_iv}
\Tr \left(\half {f^{(1,1)}}^2 + f^{(2,0)}f^{(0,2)} \right) =
\bar{\pdf}K + \pdf{\bar K}
\eeq
where
\beq\label{v_v}
\begin{array}{rcl}
K      & = & \half \Tr \left( {\bar a}\pdf a + a f^{(1,1)}
\right) \; ; \\
\bar K & = & \half \Tr \left( a \bar{\pdf}\bar a + \bar a
f^{(1,1)} \right) \; .
\end{array}
\eeq
One can easily check that $\pdf \bar{\pdf}K = \pdf \bar{\pdf}
\bar K = 0$. While eqs.~(\ref{v_v}) render the cohomology of $\pdf$, $
\bar{\pdf}$ trivial, they are not gauge invariant.

\subsec{Descent Equations and Their Solution}
\label{descent}
The fermionic symmetries $\q$, $\qq$ act as follows
\beq\label{v_vi}
\begin{array}{rclrcl}
\q a &=&  \psi - \D \omega \; ; &   {\qq} a & =& - \D {\bar{\omega}}
       \; ; \\
\q{\bar a} &=& - {\Dq} \omega  \; ; &  {\qq}{\bar a} & =&
{\bar{\psi}} - {\Dq}{\bar{\omega}} \; ; \\
\q{\psi} &=&  [ \psi , \omega ] \; ; &  {\qq}{\psi} & =& - \imu \D {\varphi} +
[ \psi , {\bar{\omega}} ] \; ; \\
\q {\bar{\psi}} &=&  - \imu {\Dq}{\varphi} + [ {\bar{\psi}} ,  \omega ] \; ; &
{\qq}{\bar{\psi}} &=& [ {\bar{\psi}} , {\bar{\omega}} ] \; ; \\
\q \omega &=& - {\omega}^2 \; ; &  {\qq}{\bar{\omega}} &=& -
{\bar{\omega}}^2 \; ;\\
\q{\varphi} &=& [ \varphi , \omega ] \; ; &  {\qq}{\varphi} &=& [ \varphi ,
{\bar{\omega}} ]\; ; \\
\imu {\varphi} &=& \q{\bar{\omega}} + \qq\omega + \{ \omega ,
{\bar{\omega}} \} \; .
\end{array}
\eeq
The ghosts $\omega$, $\bar{\omega}$ are the first components of the
Grassmann superconnections $A_{\theta}$ and $A_{\bar{\theta}}$ respectively.
They occur as supergauge transformations in superspace (see
section~\ref{wesszu}) and ensure the nilpotency and anticommutativity of the
fermionic symmetries
\beq\label{v_vii}
\q^2 = \qq^2 = \q\qq + \qq\q = 0 \; .
\eeq

Here a comment is in order, since apparently one cannot separate the action
of $\q$ on $\bar{\omega}$ from that of $\qq$ on $\omega$. In fact, the
ghosts $\omega$, $\bar{\omega}$ can be expressed through the same prepotential
$V$, as discussed in section~\ref{solut}. Of course, $V$ is determined up to
chiral gauge transformations, \ie\ up to local parameters which are annihilated
either by $\q$ or by $\qq$.

The procedure we shall now describe is an extension of the 
construction \cite{AtiyU}, \cite{Stor}, \cite{Kanno} to complex manifolds. 
Let ${\cal A}$ be 
the space of all connections on the complex vector bundle $\E$ and ${\cal G}$ 
be the group of gauge transformations. The quotient ${\cal B} = {\cal A}
{\setminus}{\cal G}$ is the set of all gauge equivalent connections. Replace
$\pdf$ by
$\Delta = \pdf + {\qq}$ and $\bar{\pdf}$ by $\bar{\Delta} = \bar{\pdf} + \q$.
The derivations $\Delta$, $\bar{\Delta}$ act over the product space 
${\cal K} \times {\cal B}$ and satisfy
\beq\label{v_viii}
{\Delta}^2 = {\bar{\Delta}}^2 = {\Delta}{\bar{\Delta}} + {\bar{\Delta}}
{\Delta} = 0 \; .
\eeq
Furthermore, one defines an adjoint bundle ${\cal E}$ over ${\cal K} \times 
{\cal B}$. Let ${\cal A} = a + {\bar{\omega}}$ and ${\bar{\cal A}} = {\bar a}
+ {\omega}$ be the connections on ${\cal E}$. One can construct generalized 
forms for the curvature
\beq\label{v_ix}
\begin{array}{c}
{\cal F}^{(2,0)} = {\Delta}{\cal A} + {\cal A}^2 \; ; \qquad
{\cal F}^{(0,2)} = {\bar{\Delta}}{\bar{\cal A}} + {\bar{\cal A}}^2
\; ; \\
{\cal F}^{(1,1)} = {\Delta}{\bar{\cal A}} + {\bar{\Delta}}{\cal A}
+ \{ {\cal A} , {\bar{\cal A}} \} \; .
\end{array}
\eeq

The quantities~(\ref{v_ix}) satisfy Bianchi identities similar to~(\ref{v_ii})
\beq\label{v_x}
\begin{array}{c}
\Delta{\cal F}^{(2,0)} + [ {\cal A}, {\cal F}^{(2,0)} ]  = 0 \; ; \qquad
\bar{\Delta}{\cal F}^{(0,2)} + [ \bar{\cal A}, {\cal F}^{(0,2)} ] = 0 \; ;
\\
\Delta{\cal F}^{(1,1)} + \bar{\Delta}{\cal F}^{(2,0)} + [ {\cal A},
{\cal F}^{(1,1)} ] + [ \bar{\cal A}, {\cal F}^{(2,0)} ] = 0 \;  ;
\\
\Delta{\cal F}^{(0,2)} + \bar{\Delta}{\cal F}^{(1,1)} + [ {\cal A},
{\cal F}^{(0,2)} ] + [ \bar{\cal A}, {\cal F}^{(1,1)} ] = 0 \; .
\end{array}
\eeq

Also basic identities look similar to~(\ref{v_iii}):
\beq\label{v_xi}
\begin{array}{rcl}
{\Delta}\; \Tr \left( \frac{1}{2} {\cal F}^{(1,1) \; 2} + {\cal F}^{(2,0)}
{\cal F}^{(0,2)} \right) + {\bar{\Delta}}\; \Tr \; {\cal F}^{(1,1)}
{\cal F}^{(2,0)} & = & 0 \; ; \\
{\bar{\Delta}}\; \Tr \left( \frac{1}{2} {\cal F}^{(1,1) \; 2} +
{\cal F}^{(2,0)}{\cal F}^{(0,2)} \right) +
{\Delta}\; \Tr \; {\cal F}^{(1,1)}{\cal F}^{(0,2)} & = & 0 \; .
\end{array}
\eeq

However, the last terms of~(\ref{v_x}) do not vanish, since
\beq\label{v_xii}
\Tr \; {\cal F}^{(1,1)} {\cal F}^{(2,0)} = \Tr f^{(2,0)} \left(
{\bar{\psi}} + \imu {\varphi} \right) \; ; \; \; \Tr \; {\cal F}^{(1,1)}
{\cal F}^{(0,2)} = \Tr f^{(0,2)} \left( {\psi} + \imu {\varphi} \right) \; .
\eeq

Due to Bianchi identities~(\ref{v_x}) the expressions~(\ref{v_xi}) are
$\Delta$\hy close and ${\bar{\Delta}}$\hy close respectively
\beq\label{v_xiii}
\Delta \Tr{\cal F}^{(1,1)}{\cal F}^{(2,0)} = 0 \; ; \qquad  \bar{\Delta}
\Tr{\cal F}^{(1,1)}{\cal F}^{(0,2)} = 0 \; .
\eeq
By enlarging the field
manifold one can make them $\Delta$ and ${\bar{\Delta}}$ exact. This feature
makes the theory of Donaldson polynomials somewhat different from that of
TYM with a  single fermionic symmetry.

Let us introduce the forms $\chi$, $b$ and $\bar{\chi}$, $\bar b$ of $(2,0)$
and $(0,2)$ type, respectively. They obey
\beq\label{v_xiv}
\begin{array}{rclrcl}
\q{\chi} & =& -\imu b - \{ \omega , \chi \} \; ; &  {\qq}{\chi} & =& -\imu
f^{(2,0)} - \{ \bar{\omega} , {\chi} \} \; ; \\
\q{\bar{\chi}} & =& - \imu f^{(0,2)} - \{ \omega , {\bar{\chi}} \} \; ; &
{\qq}{\bar{\chi}} & =& - \imu {\bar b} - \{ \bar{\omega}, {\bar{\chi}} \}
\; ; \\
\q b & =& [ b , \omega ] \; ; &  {\qq} b & =& \D {\psi} -
[ \varphi , {\chi} ] + [ b , \bar{\omega} ] \; ; \\
\q{\bar b} & =& {\Dq}{\bar{\psi}} - [ \varphi , {\bar{\chi}} ] +
[ {\bar b} , \omega ] \; ; &  {\qq}{\bar b} & =& [ {\bar b} , \bar{\omega} ]
\; .
\end{array}
\eeq
One can check that
\beq\label{v_xv}
\begin{array}{rccccl}
\Tr \; {\cal F}^{(1,1)}{\cal F}^{(2,0)} & =& \imu {\qq}\; \Tr
{\chi} \left( \bar{\psi} + \imu {\varphi} \right) & =& \imu {\Delta}\; \Tr
{\chi} \left( \bar{\psi} + \imu {\varphi} \right) &  \; ; \\
\Tr \; {\cal F}^{(1,1)}{\cal F}^{(0,2)} & =& \imu \q \; \Tr {\bar{\chi}}
\left( {\psi} + \imu {\varphi} \right) & =& \imu {\bar{\Delta}}\; 
\Tr {\bar{\chi}} \left( {\psi} + \imu {\varphi} \right) &  \; .
\end{array}
\eeq
By inserting~(\ref{v_xiv}) into~(\ref{v_x}) one gets
\beq\label{v_xvi}
\begin{array}{rcl}
\Delta \Tr \left(\half {{\cal F}^{(1,1)}}^2 + {\cal F}^{(2,0)}
{\cal F}^{(0,2)} \right) - \imu \bar{\Delta} \Tr \chi \left( \bar{\psi} +
\imu \varphi \right) & = & 0 \; ; \\
\bar{\Delta} \Tr \left(\half {{\cal F}^{(1,1)}}^2 + {\cal F}^{(2,0)}
{\cal F}^{(0,2)} \right) - \imu \Delta \Tr \bar{\chi} \left( \psi + \imu
\varphi
\right) & = & 0 \; .
\end{array}
\eeq

Let us now define
\beq\label{v_xvii}
\begin{array}{rcl}
{\cal W} & = & \ds - \frac{c}{2{\pi}^2} \left\{ \Tr \left(
\half{\cal F}^{(1,1) \; 2} + {\cal F}^{(2,0)}{\cal F}^{(0,2)} \right)
\right. \; \\
         &        & \ds   - \imu {\Delta} \left. \;
\Tr {\bar{\chi}} \left( \psi + \imu {\varphi} \right) - \imu {\bar{\Delta}}\;
\Tr {\chi} \left( {\bar{\psi}} + \imu {\varphi} \right) \right\} \; .
\end{array}
\eeq
Here the factor in front is conventional, and $c$ is the second Casimir
invariant of the adjoint representation of the ${\cal G}$, \ie\ $c_{acd}c_{bcd}
= c {\delta}_{ab}$. Eqs.~(\ref{v_xv}) take the form
\beq\label{v_xviii}
\Delta{\cal W} = 0 \; ; \qquad \bar{\Delta}{\cal W} = 0 \; .
\eeq

By expanding ${\cal W}$ in the ghost numbers associated to $\bbbr_{\pm}$
(lower bracket) one gets the descent equations:
\beq\label{v_xix}
\begin{array}{rclrcl}
\q W^{(2,2)}_{(0,0)} &\ds =& - \bar{\pdf}W^{(2,1)}_{(1,0)} \; ; & \qquad
\qq W^{(2,2)}_{(0,0)} & =& - \pdf W^{(1,2)}_{(0,1)} \; ; \\
\q W^{(2,1)}_{(1,0)} &\ds =& - \bar{\pdf}W^{(2,0)}_{(2,0)} \; ; & \qquad
\qq W^{(2,1)}_{(1,0)} & =& - \pdf W^{(1,1)}_{(1,1)} \; ; \\
\q W^{(1,2)}_{(0,1)} &\ds =& - \bar{\pdf}W^{(1,1)}_{(1,1)} \; ; & \qquad
\qq W^{(1,2)}_{(0,1)} & =& - \pdf W^{(0,2)}_{(0,2)} \; ; \\
\q W^{(1,1)}_{(1,1)} &\ds =& - \bar{\pdf}W^{(1,0)}_{(2,1)} \; ; & \qquad
\qq W^{(1,1)}_{(1,1)} & =& - \pdf W^{(0,1)}_{(2,1)} \; ; \\
\q W^{(2,0)}_{(2,0)} &\ds =& 0 \; ;                                 & \qquad
\qq W^{(2,0)}_{(2,0)} & =& - \pdf W^{(1,0)}_{(2,1)} \; ; \\
\q W^{(0,2)}_{(0,2)} &\ds =& - \bar{\pdf} W^{(0,1)}_{(1,2)} \; ; & \qquad
\qq W^{(0,2)}_{(0,2)} & =& 0 \; ;  \\
\q W^{(1,0)}_{(1,2)} &\ds =& 0 \; ;                                  & \qquad
\qq W^{(1,0)}_{(2,1)} & =& - \pdf W^{(0,0)}_{(2,2)} \; ; \\
\q W^{(0,1)}_{(1,2)} &\ds =& - \bar{\pdf} W^{(0,0)}_{(2,2)} \; ; & \qquad
\qq W^{(0,1)}_{(1,2)} & =& 0 \; ;  \\
\q W^{(0,0)}_{(0,0)} &\ds =& 0 \; ;                                  & \qquad
\qq W^{(0,0)}_{(2,2)} & =& 0 \; .
\end{array}\eeq
The upper bracket indicates the type of form on K\"{a}hler manifold.

In order to obtain the solution of~(\ref{v_xix}) we write~(\ref{v_xvii}) in
the form
\beq\label{v_xx}
\begin{array}{rcl}
{\cal W} & = & \ds - \frac{c}{4{\pi}^2}\; \Tr \; { \left( f^{(1,1)} +
f^{(2,0)} + f^{(0,2)} + {\psi} + {\bar{\psi}} + \imu {\varphi} \right) }^2 \;
\\
         &   & \ds +  \frac{\imu c}{2{\pi}^2} \left\{ {\Delta}\;
\Tr {\bar{\chi}} \left( \psi + \imu {\varphi} \right) + {\bar{\Delta}}\;
\Tr {\chi} \left( {\bar{\psi}} + \imu {\varphi} \right) \right\} \; .
\end{array}
\eeq

By expanding ${\cal W}$ according to $r_{\pm}$ numbers one gets
\beq\label{v_xxi}
\vcenter{\hbox{%
$\begin{array}{rcl}
W^{(2,2)}_{(0,0)} & = &\ds \frac{c}{4{\pi}^2} \Tr \left\{\half {f^{(1,1)}}^2
+ f^{(2,0)}f^{(0,2)} - \imu \pdf \left( \bar{\chi}\psi \right) -
\imu \bar{\pdf}\left( \chi \bar{\psi}\right)
\right\} \; ; \\
W^{(2,1)}_{(1,0)} & = &\ds \frac{c}{4{\pi}^2} \Tr \left( \varphi {\Dq}\chi
- f^{(1,1)}\psi - b \bar{\psi}  \right)\; ; \\
W^{(1,2)}_{(0,1)} & = &\ds \frac{c}{4{\pi}^2} \Tr \left( \varphi \D \bar{\chi}
- f^{(1,1)}\bar{\psi} - \bar{b}\psi \right) \; ; \\
\end{array}$}\hbox{%
$\begin{array}{rclrcl}
W^{(2,0)}_{(2,0)} & = &\ds \frac{c}{4{\pi}^2} \Tr \left(\half\psi^2 - \imu
\varphi b  \right) \; ;\;& W^{(0,2)}_{(0,2)} &=& \ds\frac{c}{4{\pi}^2}
\Tr \left(\half\bar{\psi}^2 - \imu \varphi \bar{b}  \right) \; ;
\\
W^{(1,1)}_{(1,1)} & = &\ds \frac{c}{4{\pi}^2} \Tr \left( \imu \varphi f^{(1,1)}
+
\psi \bar{\psi} \right) \; ;\;& W^{(1,0)}_{(2,1)} &=&\ds
\frac{\imu c}{4{\pi}^2} \Tr \varphi \psi \; ; \\
W^{(0,1)}_{(1,2)} & = &\ds \frac{\imu c}{4{\pi}^2} \Tr \varphi \bar{\psi} \;
;\;&
W^{(0,0)}_{(2,2)} &=& \ds - \frac{c}{8{\pi}^2} \Tr {\varphi}^2 \; .
\end{array}$}}
\eeq
The  numbers in brackets can be checked by using their additivity as well as
table~1.
\begin{table}\label{tab}
$$
\begin{array} {||l|c|c|c|c|r|r|r|r|r|r|r|r|c||} \hline
        & a & \bar{a} & \psi &\vphantom{\bar{\bar{\psi}}}\bar{\psi} & \chi &
\bar{\chi} & b  & \bar{b} & \varphi & \lambda & g_{+} & g_{-} & k \\ \hline
r_+       & 0 & 0       & 1    & 0          & 0    & -1         & 1  & -1
&
1       & -1      & 0     & -1    & 0 \\ \hline
r_-       & 0 & 0       & 0    & 1          & -1   & 0          & -1 & 1
&
1       & -1      & -1    & 0     & 0 \\ \hline
p       & 1 & 0       & 1    & 0          & 2    & 0          & 2  & 0       &
0       & 0       & 0     & 0     & 0 \\ \hline
q       & 0 & 1       & 0    & 1          & 0    & 2          & 0  & 2       &
0       & 0       & 0     & 0     & 0 \\ \hline
d       & 1 & 1       & 1    & 1          & 2    & 2          & 2  & 2       &
0       & 2       & 2     & 2     & 2 \\ \hline
\end{array}
$$
\caption{Quantum numbers and form degrees of various fields.}
\end{table}
The meaning of the letters in the first column is the following: $r_+$, $r_-$
are the quantum numbers of the global Abelian symmetries,
$(p, q)$ is the complex form degree and $d$ the canonical dimension of the 
field.

One can show further that
\beq\label{v_xxii}
W^{(0,0)}_{(2,2)} = \frac{c}{8{\pi}^2} \left\{ \q \Tr \left( \imu \varphi -
\omega \bar{\omega} \right) \bar{\omega} + \qq \Tr \left( \imu
\varphi + \bar{\omega} \omega \right) \omega \right\} \; .
\eeq
Nevertheless, $W^{(0,0)}_{(2,2)}$ is a nontrivial element of the (equivariant)
cohomology of $\q$ and $\qq$, since it does not depend on $\omega$ or
$\bar{\omega}$. In other words, both $\Tr ( \imu \varphi - \omega
\bar{\omega} ) \bar{\omega}$ and $\Tr ( \imu \varphi + \bar{\omega}\omega
) \omega$ are not gauge invariant.

Hence, $W^{(p,q)}_{(r_+,r_-)}$ are local
observables whose correlation functions might be nonvanishing.
Of course, the $W^{(p,q)}_{(r_+,r_-)}$ are gauge invariant, \ie\ $\s
W^{(p,q)}_{(r_+,r_-)} = 0$. Notice that $\s$, $\q$ and $\qq$  anticommute 
with each other.

\subsec{Cohomology Classes}
\label{cohom}
Let us consider the equivalence classes of $(p,q)$\hy forms which are both
$\pdf$ and $\bar{\pdf}$ closed but not exact. A $(p,q)$\hy form is
exact if either ${\omega}_{(p,q)} = \pdf \bar{\pdf}{\phi}_{(p-1,q-1)}$
or ${\omega}_{(p,0)} = \pdf {\phi}_{p-1}(z)$ or else ${\omega}_{(0,q)} =
\bar{\pdf}{\bar{\phi}}_{q-1}(\bar z)$, respectively; here $p, q {\geq}1$.
The equivalence classes make up a vector space known as the Dolbeault
cohomology group ${\cal H}^{(p,q)}({\cal K}; \pdf, \bar{\pdf} )$.
(The composition law is the additive group structure of the vector space.)
Let us define
\beq\label{v_xxiii}
{\Omega}_{(r_+,r_-)} = \int_{{\cal K}}W^{(2-p,2-q)}_{(r_+,r_-)}{\omega}_{(p,q)}
\eeq
where ${\omega}_{(p,q)}$ is a $\pdf$ and $\bar{\pdf}$ closed $(p,q)$\hy
form independent of the fields. One can check that~(\ref{v_xxiii}) is
annihilated by both fermionic charges $\q$, $\qq$. Furthermore, if
${\omega}_{(p,q)}$ is exact, then ${\Omega}_{(r_+=q,r_-=p)}$ is highest
component, \ie\ it can be written in one of the following ways: $\q \qq
{\Phi}_{(q-1,p-1)}$, $\q{\Phi}_{(q-1,0)}$ or $\qq {\Phi}_{(0,p-1)}$.

We shall see in the next subsection that $\q$, $\qq$ can be interpreted
as complex exterior derivatives on the instanton moduli space ${\cal M}$.
Since ${\Omega}_{(q,p)}$ is a $(q,p)$\hy form closed with respect to both $\q$
and $\qq$, it belongs to the Dolbeault group ${\cal H}^{(q,p)}({\cal M}; \q,
\qq )$. The Donaldson map between the Dolbeault cohomology groups relates
${\omega}_{(p,q)}\in {\cal H}^{(p,q)}({\cal K}; \pdf, \bar{\pdf} )$
to ${\Omega}_{(q,p)}\in {\cal H}^{(q,p)}({\cal M}; \q, \qq )$.

\subsec{Donaldson Invariants}
\label{doninv}
For computing invariant correlation functions one needs the integrated
observables~(\ref{v_xxiii}). It is convenient to express  them in the form:
\begin{eqnarray}\label{v_xxiv}
{\Omega}_{(0,0)} & = & \frac{c}{4{\pi}^2} \int_{\cal K} \Tr \left( \half
{f^{(1,1)}}^2 + f^{(2,0)}f^{(0,2)} \right) \; ; \nonumber \\
{\Omega}_{(1,0)} & = & - \frac{c}{4{\pi}^2} \int_{\cal K} \Tr \left\{
f^{(1,1)}\psi + \imu \q \left( \chi \bar{\psi} \right) \right\}
{\omega}_{(0,1)}
\; ; \nonumber \\
{\Omega}_{(0,1)} & = & - \frac{c}{4{\pi}^2} \int_{\cal K} \Tr \left\{
f^{(1,1)}\bar{\psi} + \imu \qq \left( \bar{\chi} \psi \right) \right\}
{\omega}_{(1,0)} \; ; \nonumber \\
{\Omega}_{(2,0)} & = & \frac{c}{4{\pi}^2} \int_{\cal K} \Tr \left\{
\half {\psi}^2 + \q \left( \chi \varphi \right) \right\} {\omega}_{(0,2)}
\; ;  \nonumber \\
{\Omega}_{(0,2)} & = & \frac{c}{4{\pi}^2} \int_{\cal K} \Tr \left\{
\half{\bar{\psi}}^2 + \qq \left( \bar{\chi} \varphi \right) \right\}
{\omega}_{(2,0)} \; ; \\
{\Omega}_{(1,1)} & = & \frac{c}{4{\pi}^2} \int_{\cal K} \Tr \left( \imu \varphi
f^{(1,1)} +  \psi \bar{\psi} \right) {\omega}_{(1,1)} \; ; \nonumber \\
{\Omega}_{(2,1)} & = & \frac{\imu c}{4{\pi}^2} \int_{\cal K} \Tr \varphi \psi
{\omega}_{(1,2)} \; ;  \qquad {\Omega}_{(1,2)}  = \frac{\imu c}{4{\pi}^2}
\int_{\cal K} \Tr \varphi \bar{\psi} {\omega}_{(2,1)} \; ; \nonumber \\
{\Omega}_{(2,2)} & = & - \frac{c}{8{\pi}^2} \int_{\cal K} \Tr {\varphi}^2
{\omega}_{(2,2)} \; . \nonumber
\end{eqnarray}

The correlation functions of the observables~(\ref{v_xxiv}) are the well-known 
Donaldson invariants and have the form
\beq\label{v_xxv}
\vev{\prod_{i} {\Omega}_{(p_i,q_i)}} = \int [\md\mu] \prod_{i}
{\Omega}_{(p_i,q_i)} \,
\exp \left\{ - \frac{1}{e^2} {\cal S} \right\} \; .
\eeq
where the functional integral is performed upon the fields $\mu$ figuring 
in table~1.
Since the observables ${\Omega}_{(p,0)}$, ${\Omega}_{(0,q)}$ depend on
$\chi$, $\bar{\chi}$, a direct integration of the non\hy zero modes in the 
path integral is not possible.
Nevertheless, one can show that the correlation 
functions~(\ref{v_xxv}) remain unchanged if the equations~(\ref{v_xxiv}) 
are replaced by 
another system of observables depending 
only upon the gauge fields and their various topological ghosts. The new  
observables -- denoted by ${\widetilde{\Omega}}_{(p,q)}$ -- are obtained from 
${\Omega}_{(p,q)}$ by interchanging $\q$ and $\qq$. A simple calculation 
leads to 
\begin{eqnarray}\label{v_xxvi}
{\widetilde{\Omega}}_{(1,0)} & = & - \frac{c}{4{\pi}^2} \int_{\cal K} 
\Tr \left\{
f^{(1,1)}\psi + f^{(2,0)} \bar{\psi} \right\}{\omega}_{(0,1)}
\; ; \nonumber \\
{\widetilde{\Omega}}_{(0,1)} & = & - \frac{c}{4{\pi}^2} \int_{\cal K} 
\Tr \left\{
f^{(1,1)}\bar{\psi} + f^{(0,2)} \psi \right\}
{\omega}_{(1,0)} \; ;   \\
{\widetilde{\Omega}}_{(2,0)} & = & \frac{c}{4{\pi}^2} \int_{\cal K} 
\Tr \left\{
\half {\psi}^2 - \imu f^{(2,0)} \varphi \right\} {\omega}_{(0,2)}
\; ;  \nonumber \\
{\widetilde{\Omega}}_{(0,2)} & = & \frac{c}{4{\pi}^2} \int_{\cal K} 
\Tr \left\{
\half{\bar{\psi}}^2 - \imu f^{(0,2)} \varphi \right\}
{\omega}_{(2,0)} \; , \nonumber
\end{eqnarray}
while the other observables obviously are not affected.

For the proof let us write the generating functional of~(\ref{v_xxv}) in the 
form (${\alpha}_q$, ${\beta}_p$ are arbitrary numbers)
\beq\label{v_xxvii}
\vev{ \exp \left\{ \sum_q {\alpha}_q \left( A_q + \qq B_q \right) + 
\sum_p {\beta}_p \left( C_p + \q D_p \right) \right\} } \; 
\eeq
where  $A_q + \qq B_q$, 
$C_p + \q D_p$ are the Donaldson integrated observables~(\ref{v_xxiv}). Being 
invariant under both $\q$ and $\qq$ they obey 
\beq\label{v_xxviii}
\qq A_q = \q C_p = 0 \; ; \quad \q A_q = \qq \q B_q \; ; \quad \qq
C_p = \q \qq D_p \; .
\eeq
Of course, for some $q$ or $p$ one can have $B_q = 0$ or $D_p = 0$.

Let us deform~(\ref{v_xxvii}) continuosly into the generating functional of
the new observables ${\widetilde{\Omega}}_{(p,q)}$ by means of
\beq\label{v_xxix}
\vev{ \exp \left\{ \sum_q {\alpha}_q \left[ A_q + \left( (1 - u)\qq + u \q 
\right) B_q \right] + 
\sum_p {\beta}_p \left[ C_p + \left( (1 - u)\q + u \qq \right) D_p \right] 
\right\} } \; 
\eeq
where $u$ is a real parameter in the interval $[ 0, 1 ]$. By differentiating
with respect to $u$ and by using the conditions~(\ref{v_xxviii}) one can show
that~(\ref{v_xxix}) does not depend on $u$. The generating functional of the
new observables is obtained from~(\ref{v_xxix}) by taking $u = 1$. Due to the
$u$ independence it coincides, however, with~(\ref{v_xxvii}). The integrand of
any correlation function has been transformed into an expression depending
only on the variables $a$, $\bar a$, their topological ghosts $\psi$,
$\bar{\psi}$ and the ghost for ghosts $\varphi$. Hence we proved
\beq\label{v_xxx}
\vev{\prod_{i} {\Omega}_{(p_i,q_i)}} = \vev{\prod_{i} 
{\widetilde{\Omega}}_{(p_i,q_i)}} \; .
\eeq

The evaluation of the correlation function proceeds along the same lines as
for TYM with a single fermionic symmetry \cite{I}, by taking advantage of 
working with the action
\beq\label{v_xxxi}
\frac{1}{8} \q \qq \int_{\cal K} \Tr \left( \chi \bar{\chi} - \imu {\gamma}^2 
\lambda f \right) \; 
\eeq
which differs from (2.10) by a $\q \qq $\hy term.  
One can now integrate out all non\hy zero modes. It is usually assumed that
there are no zero modes in the variables $\chi$, $\bar{\chi}$, $b$ and $\bar
b$.  

There is a
special prescription for handling the ghost for ghosts: The field $\varphi$ has
to be replaced by the solution $\vev{\varphi}$ of the differential equation
\cite{Park}
\beq\label{v_xxxii}
g^{\bar n m} \left( \{ \D_m , \D_{\bar n} \} \vev{\varphi} + 2 \imu \{ {\psi}_m
, {\psi}_{\bar n} \} \right) = 0 \;
\eeq
where ${\psi}_m$, ${\psi}_{\bar m}$ are the zero modes of the topological
ghosts. 
In solving eq.~(\ref{v_xxxii}) one can meet zero modes of $\varphi$,
for which the procedure of ref.~\cite{Anse} should be extended to the
K\"{a}hler case. For simplicity we shall assume in the following that 
also such zero modes are absent.

The path integral measure takes its canonical form $[\md a] [\md \bar a] [\md
\psi] [\md\bar{\psi}]$ where $a$ and $\bar a$ are solutions of the self\hy
duality conditions:
\beq\label{v_xxxiii}
f_{mn} = f_{\bar m \bar n} = g^{\bar n m} f_{m \bar n} = 0 \; .
\eeq
Since $\psi$ and $\bar{\psi}$ are the zero modes of the topological ghosts,
they obey the following equations of motion:
\beq\label{v_xxxiv}
\D_{[m} {\psi}_{n]} = \D_{[ \bar m } {\psi}_{\bar n ]} = \D_m {\psi}^m = \D^m
{\psi}_m = 0 \; .
\eeq
One can show \cite{I}, \cite{Park} that instanton deformations orthogonal to
purely
gauge transformations obey identical equations.
Hence $\psi$ and $\bar{\psi}$ are tangent vectors to the instanton moduli
space ${\cal M}$. Since $\q$, $\qq$ relate $a$, $\bar a$ to $\psi$,
$\bar{\psi}$, they play the role of exterior derivatives on ${\cal M}$.

The integration of $\psi$, $\bar{\psi}$ is straightforward and transforms the
integrand into a wedge product of $(p_i,q_i )$\hy forms over the moduli space
\beq\label{v_xxxv}
\vev{\prod_{i} {\Omega}_{(p_i,q_i)} } = \int_{\cal M} \prod_{i}
{\Phi}_{(p_i,q_i)} \; .
\eeq
In writing down eq.~(\ref{v_xxxv}) we assumed that ${\cal M}$ can be
considered a finite dimensional K\"{a}hler manifold \cite{Koba}.

One can now establish a selection rule for the correlation functions as given
by~(\ref{v_xxxv}). The action ${\cal S}$ needed
for computing the left hand side is invariant under the global Abelian symmetry
$\bbbr_{+}\otimes \bbbr_{-}$. In contrast the integration measure transforms
under $\bbbr_{\pm}$ with certain weights that are equal and exactly compensate
the dimension of ${\cal M}$. Therefore ${\Omega}_{(p_i,q_i)}$ should provide
the compensating total weights
\beq\label{v_xxxvi}
\sum_{i} p_i = \sum_{i} q_i = \dim {\cal M} \; .
\eeq
This means that the integrand of~(\ref{v_xxxv}) is a top\hy form, \ie\ a $(
\dim {\cal M}, \dim {\cal M} )$\hy form over ${\cal M}$.

The careful reader may have noticed that the correlation functions were defined 
by using the action (2.10) in which the BRS gauge fixing has been neglected. 
The importance of the BRS gauge fixing conditions both for interpreting and 
computing Donaldson invariants has been emphasized for TYM with a single 
fermionic charge in \cite{XI}. 

In our case the BRS gauge fixing appears in the total action $\widetilde{\cal
S} + {\cal S}'$ suggesting its use in defining the correlation functions.
Since the eqs.~(\ref{v_xiv}) relating $\chi$, $\bar{\chi}$ to $f^{(2,0)}$ and
$f^{(0,2)}$ respectively, now become equations of motion, we cannot start from
the old observables ${\Omega}_{(p,q)}$, but rather from the new ones
${\widetilde{\Omega}}_{(p,q)}$. After performing the functional integration
over the chiral superfields $M_{mn}$, $M_{\bar m \bar n}$ one recovers the
full system of eqs.~(\ref{v_xiv}) and one can infer that~(\ref{v_xxx}) still
holds (albeit with a gauge fixed action). This is important in order to make
sure that we are discussing the correlation functions of the solution to the
cohomology problem of $\q$ and $\qq$.

It is possible to develop an analysis for TYM with two fermionic charges
similar to that performed in \cite{XI} in order to show that the BRS gauge
fixing in superspace is equivalent to Witten's method of computing the
correlation functions. Let us write the gauge\hy fixing action in the form
\beq\label{v_xxxvii}
-\quart\,\s\int_{\cal K}\gamma\left\{\q\,\Tr\, d\bar\pdf u
  - \qq\,\Tr\, d^\kr\pdf\bar u\right\}\;.
\eeq
Here $d$ and $d^\kr$ are the first components of the chiral superfields $D$
and $D^\kr$. The $(1,0)$ and $(0,1)$ forms
\beq\label{v_xxxviii}
u = u_m \md z^m\;;\qquad \bar u = u_{\bar m}\md z^{\bar m}
\eeq
are constructed from the chiral connection superfields
\beq\label{v_xxxix}
\phi_m = u_m + \theta\pi_m\;;\qquad \phi_{\bar m} = u_{\bar m} + 
\bar{\theta}{\pi}_{\bar m} \: .
\eeq
In view of the above gauge fixing term one can start from the following path
integral for the correlation functions
\beq\label{v_xl}
\vev{{\cal O}} = \int [\md \hat\mu][\md V]\,{\cal O}(\hat\mu,V)\,
                 \exp{\left\{-\frac{1}{e^2}{\cal S}[\hat\mu,V]\right\}}
\delta(\nabla^m\phi_m)
                                          \delta(\nabla^{\bar m}\phi_{\bar m})
                                      \hat\Delta(\phi,\bar\phi)\;
\eeq
where $\hat\mu$ represents the collection of superfields $\phi_m$, $\phi_{\bar
m}$, $\Lambda$, $X_{mn}$, $X_{\bar m\bar n}$; ${\cal O}(\hat\mu,V)$ denotes a
gauge invariant function (a product of Donaldson polynomials) and
\beq\label{v_xli}
\hat\Delta^{-1}(\phi,\bar\phi)
  = \int_{\cal G}[\md g]\,\delta(\nabla^m\phi_m^g)
                          \delta(\nabla^{\bar m}\phi_{\bar m}^g)
\eeq
is the Faddeev\Et Popov (super)determinant.

From now on we will express all superfields by components. We would like to
consider the gauge group ${\cal G}$ consisting of chiral transformations with
the parameters $\eta$ and $\eta^\kr$ acting on the field components as follows:
\beq\label{v_xlii}
\begin{array}{rclrcl}
u'_m &=& \e^{-\eta} (u_m + \nabla_m) \e^\eta\;;\quad&
u'_{\bar m} &=& \e^{\eta^\kr} (u_{\bar m} + \nabla_{\bar m})
                    \e^{-\eta^\kr}\;;\\
\pi'_m &=& \e^{-\eta} \pi_m \e^\eta\;;\quad&
\pi'_{\bar m} &=& \e^{\eta^\kr} \pi_{\bar m}
                    \e^{-\eta^\kr}\;;\\
\e^{v'} &=& \e^{\eta\kr} \e^v \e^\eta\;
\end{array}
\eeq
where $v$ is the first component of the superfield $V$. The components of 
non\hy chiral gauge superfields transform according to unitary transformations
generated by $h^\kr=-h$
\beq\label{v_xliii}
\begin{array}{rclrcl}
a'_m &=& \e^{-h} (a_m + \nabla_m) \e^h\;;\quad&
a'_{\bar m} &=& \e^{-h} (a_{\bar m} + \nabla_{\bar m})
                    \e^h\;;\\
\pi'_m &=& \e^{-h} \pi_m \e^h\;;\quad&
\pi'_{\bar m} &=& \e^{-h} \pi_{\bar m}
                    \e^h\;;\\
\omega' &=& \e^{-h} \omega \e^h\;;\quad&
\bar\omega' &=& \e^{-h} \bar\omega \e^h\;;\\
\varphi' &=& \e^{-h}\varphi\e^h\;.
\end{array}
\eeq
(As can be seen from (2.17) $h$ depends highly non\hy trivially upon $\eta$,
$\eta^\kr$, and $v$. The components of $\Lambda$, $X_{mn}$, and 
$X_{\bar m\bar n}$ transform similarly, but they are not of interest for us
here.)
 Finally, the matrix
$\e^{\frac{v}{2}}$ transforms in one of the following equivalent ways
\beq\label{v_xliv}
\e^{\frac{v'}{2}} = \e^{\eta^\kr}\e^{\frac{v}{2}}\e^h
                  = \e^{-h}\e^{\frac{v}{2}}\e^{\eta}
\eeq
and serves to relate components of chiral and of gauge superfields
\beq\label{v_xlv}
\begin{array}{rclrcl}
u_m &=& \e^{-\frac{v}{2}} (a_m + \nabla_m) \e^{\frac{v}{2}}\;;\quad&
u_{\bar m} &=& \e^{\frac{v}{2}} (a_{\bar m} + \nabla_{\bar m})
                    \e^{-\frac{v}{2}}\;;\\
\pi_m &=& {\cal D}_m\omega
  + \e^{-\frac{v}{2}}\left(\psi_m - \D_m\omega\right)\e^{\frac{v}{2}}\;;\quad&
\pi_{\bar m} &=& {\cal D}_{\bar m}\bar\omega
  + \e^{\frac{v}{2}}\left(\psi_{\bar m} - \D_{\bar m}\bar\omega\right)
                           \e^{-\frac{v}{2}}\;.
\end{array}
\eeq

Let us now perform a chiral transformation on the path integral measure and on
the integrand of~(\ref{v_xl}). Everything but the $\delta$\hy function is
invariant under such a transformation.

If one chooses the chiral transformation such that $v'=0$, one gets from
~(\ref{v_xlii})--(\ref{v_xliv}) the relations
\beq\label{v_xlvi}
\begin{array}{rclrcl}
\e^\eta &=& \e^{-\frac{v}{2}}\e^h\;;\quad&
\e^{\eta^\kr} &=& \e^{-h}\e^{-\frac{v}{2}}\;;\\
u'_m &=& a'_m\;;\quad&
u'_{\bar m} &=& a'_{\bar m}\;;\\
\pi'_m &=& \psi'_m\;;\quad&
\pi'_{\bar m} &=& \psi'_{\bar m}\;.
\end{array}
\eeq
One can show  that
\begin{eqnarray}\label{v_lii}
V        & = & 2 \theta {\omega}' - 2 \bar{\theta}{\bar{\omega}}' + \theta 
\bar{\theta} \left( \imu {\varphi}' - 2 \{ {\omega}', 
{\bar{\omega}}' \} \right) \; ; \nonumber \\
{\phi}_m & = & {a'}_m + \theta{{\psi}'}_m \; ; \qquad {\phi}_{\bar m} = 
{a'}_{\bar m} + \bar{\theta}{{\psi}'}_{\bar m} \;  
\end{eqnarray}  
is the supersymmetry gauge in which $\q = {\partial}_{\theta}$ and $\qq = 
{\partial}_{\bar{\theta}}$ have the action given by~(\ref{v_vi}). 

Now we perform the change of field variables
\beq\label{v_xlvii}
\begin{array}{rclrcl}
u_m &\to& a_m\;;\quad&
u_{\bar m} &\to& a_{\bar m}\;;\\
\pi_m &\to& \psi_m\;;\quad&
\pi_{\bar m} &\to& \psi_{\bar m}\;,
\end{array}
\eeq
so that any dependence of $v$, $\omega$, and
$\bar\omega$ in ${\cal S}$ and ${\cal O}$ disappears and moreover the 
corresponding Jacobians are equal to one. The system of variables
$\hat\mu$ is replaced by $\mu$. One can easily check that 
\beq\label{v_xlviii}
\begin{array}{rl}
\ds\vev{{\cal O}} =  \int [\md\mu] & \ds {\cal O}(\mu)\,\exp{\left\{-
\frac{1}{e^2}{\cal S}[\mu]\right\}} \\
 &\ds \times \delta(\nabla^m a'_m)
    \delta(\nabla^{\bar m} a'_{\bar m})
    \delta(\nabla^m \psi'_m)
    \delta(\nabla^{\bar m} \psi'_{\bar m})
    \hat{\Delta}(a,\psi;\bar a,\bar\psi)\;
\end{array}
\eeq
where the new Faddeev\Et Popov (super)determinant is obtained by integrating
over the unitary subgroup generated by $h$:
\beq\label{v_xlix}
\begin{array}{rcl}
\ds\hat{\Delta}^{-1}(a,\psi;\bar a,\bar\psi) &=& \ds \int [\md h]\,
    \delta(\nabla^m a'_m)
    \delta(\nabla^{\bar m} a'_{\bar m}) \\
&&\ds \times \delta(\nabla^m \psi'_m)
    \delta(\nabla^{\bar m} \psi'_{\bar m}) \; .
\end{array}
\eeq

The above considerations lead to 
the following path integral for the correlation functions:
\beq\label{v_l}
\vev{{\cal O}} = \int [\md\mu][\md\nu]\, {\cal O}(\mu)\, 
\exp{\left\{-\frac{1}{e^2}{\cal S}[\mu,\nu]\right\}}\;
\eeq
where
\beq\label{v_li}
\begin{array}{rcl}
{\cal S}[\mu,\nu] &=&\ds\quart\int_{\cal K}\Bigl\{\half\q\qq\,\Tr \left( 
\chi \bar{\chi} -  \imu {\gamma}^2 \lambda f \right)\Bigr.\\
          &&\ds\Bigl. -\,\s \gamma \left[\q\,\Tr\, d\, \Dq_0(a-a_0) - \qq\,\Tr\,
               d^\kr\,\D_0(\bar a - \bar a_0)\right]\Bigr\}\;
\end{array}
\eeq
and $\nu$ stands for the fields $d$, $d^\kr$ as well as all the fields obtained
from them by applying $\s$, $\q$, and $\qq$. We also introduced the background
gauge field forms $a_0$, $\bar a_0$  and the corresponding covariant 
differentials $\D_0$, $\Dq_0$ upon which $\s$, $\q$, and $\qq$ act trivially.

The similarity of~(\ref{v_l}) with the expression used in~\cite{XI} shows that
the prescription to evaluate the Donaldson invariants can be derived from the
standard (with gauge fixed action) path integral also for TYM with two
fermionic charges. Of course, the prescription coincides with that obtained by
neglecting the gauge fixing term.

A first systematic attempt to compute Donaldson invariants of smooth, oriented, 
compact four\hy manifolds has been given by Kronheimer and Mrowka \cite{KM}. 
They showed that the Donaldson invariants of the so\hy called manifols of 
simple type 
exhibit universal relations. It has been conjectured \cite{KM}, \cite{Mo} that 
all simply\hy connected four\hy manifolds with $b_2^{+}$ ( $b_2^{+}$ 
is the number of independent self\hy dual forms ) are of simple type. 
Subsequently, almost all $SU(2)$ 
and $SO(3)$ Donaldson invariants for K\"ahler four\hy manifolds of 
simple type with 
$b_2^{+} = \dim {\cal H}^{(1,1)}({\cal K};\partial ,\bar{\partial}) > 1$ 
have been calculated by Witten \cite{III}, making use of the known infrared 
behaviour of $N=1$ supersymmetric gauge theories. On the other hand, 
precise formulas relating the Donaldson invariants to Seiberg\Et Witten 
invariants ( for a review see \cite{mm} ) have been conjectured in \cite{Mo}. 
In sharp contrast to Donaldson 
invariants, which are defined on the moduli space of instantons, Seiberg\Et 
Witten invariants are associated to moduli spaces of abelian monopoles. 
The Seiberg\Et Witten theory is a powerful method which allows   
the calculation of all Donaldson invariants in case of K\"ahler manifolds of 
simple type as mentioned above.
In order to make contact with 
the present work we point out that the manifolds of simple type can only 
have correlation functions of the observables ${\Omega}_{(1,1)}$ and 
${\Omega}_{(2,2)}$. 

A different approach based on the holomorphic Yang\Et Mills theory \cite{Holo} 
has been proposed in \cite{Hyun} and used for computing correlation 
functions of the product ${\Omega}_{(2,0)}{\Omega}_{(0,2)}$.
 
Concerning the mathematical literature we refer to \cite{FS} where 
the Donaldson invariants are obtained by means of the so-called blowup 
formula. A first step in proving the formulas conjectured by Witten \cite{Mo} 
has been made in \cite{PT}.

Finally let us mention two papers \cite{HP}, \cite{Sd} where the question of 
computing Donaldson invariants for four\hy manifolds with $b_2^{+} \leq 1$ 
is raised.
 
Some results of this section have been obtained in refs.\ \cite{Park},
\cite{Holo}. They concern Donaldson observables with equal ghost numbers
${\Omega}_{(p,p)}$.  Here we included the off\hy diagonal Donaldson
observables, thereby completing the interpretation of the fermionic charges as
complex derivations on the instanton moduli space.

\sect{Conclusions}

In the present paper we formulated TYM theory with two fermionic charges on the
superspace consisting of a K\"{a}hler four\hy manifold and two Grassmann
variables. In contrast to TYM theory with a single fermionic charge, we had to
impose certain constraints in superspace. We solved the constraints and showed
that the gauge transformations were replaced by local chiral transformations.
Then we elucidated the structure of the Faddeev\Et Popov ghost sector and 
determined
the total action.

Furthermore, we used the action for perturbatively computing 
the (K\"{a}hler) gravitational contribution to the dimension of the instanton
moduli space. In performing the calculation we showed how the covariant
point\hy splitting technique can be extended to K\"{a}hler manifolds.

Insisting on
the specific form taken by the local conservation law on K\"{a}hler manifolds 
we discussed in some detail the global symmetries of the action. 
We showed that the associated currents representing locally these symmetries
(en\-er\-gy\hy momentum tensor, fermionic and antisymmetry currents)
are highest components of gauge invariant superfields. BRS symmetry does not
alter this property, while the irreducibility of the multiplets is
sometime lost. In any case, all their correlation functions vanish.

In addition we also determined the non\hy trivial observables. 
They are cohomology classes
of both fermionic symmetry operations. Some of the classes involve additional
fields, absent in TYM with a single fermionic charge. Nevertheless, we could 
show that the correlation functions of all non\hy trivial observables can be 
represented as integrals of top\hy forms over the instanton moduli space.

\section*{Acknowledgements}
We would like to thank Friedemann Brandt for his comments on the
algebraic cohomology problem. One of us (S.\ M.) would like to acknowledge
useful discussions with Oleg Ogievetsky and Raymond Stora.

\end{document}